\documentclass[12pt]{article}
\usepackage{graphicx,epsfig}
\input axodraw.sty
\usepackage{amsfonts,amssymb,amsmath,amssymb}
\usepackage[dvips]{color}
\newcommand{\ba}{\begin{array}}
\newcommand{\ea}{\end{array}}
\newcommand{\beq}{\begin{equation}}
\newcommand{\eeq}{\end{equation}}
\def\bt{\begin{table}}
\def\et{\end{table}}
\def\bc{\begin{center}}
\def\ec{\end{center}}
\def\bi{\begin{itemize}}
\def\ei{\end{itemize}}
\def\bea{\begin{eqnarray}}
\def\eea{\end{eqnarray}}
\def\beas{\begin{eqnarray*}}
\def\eeas{\end{eqnarray*}}

\def\N0{\widetilde{\chi}^0}

\catcode`@=11 
\def \gsim{\mathrel{\mathpalette\@versim>}}
\def \lsim{\mathrel{\mathpalette\@versim<}}
\def \@versim#1#2{\lower0.4ex\vbox{\baselineskip\z@skip\lineskip\z@skip
     \lineskiplimit\z@\ialign{$\m@th#1\hfil##\hfil$%
     \crcr#2\crcr\sim\crcr}}} 
\catcode`@=12 

\begin{document} 
\setcounter{page}{0} 
\thispagestyle{empty} 
\begin{flushright} HIP-2010-33/TH \\
                   OSU-HEP-10-09  \end{flushright} 
\begin{center}
{\Large  Probing top charged-Higgs production using top polarization at the
Large Hadron Collider} 
\\ \vspace*{0.2in} 
{\large  Katri Huitu$^1$, Santosh Kumar Rai$^2$, Kumar Rao$^1$, Saurabh D.
Rindani$^3$ and Pankaj Sharma$^3$} \\
\vspace*{0.2in}
{\sl $^1$Department of Physics, University of Helsinki, and
                                 Helsinki Institute of Physics, \\
          P.O. Box 64, FIN-00014 Helsinki, Finland\\
     $^2$ Department of Physics 
              and Oklahoma Center for High Energy Physics, \\
              Oklahoma State University, 
              Stillwater, OK 74078, USA \\
     $^3$ Theoretical Physics Division, Physical Research Laboratory, \\
          Navrangpura, Ahmedabad 380 009, India
\rm }
\end{center}
\vspace*{0.6in}
\begin{center}
{\large\bf Abstract}
\end{center}

We study single top production in association with a charged Higgs in the
type II two Higgs doublet model at the Large Hadron Collider. The polarization
of the top, reflected in the angular distributions of its decay products, can be
a sensitive probe of new physics in its production. We present theoretically
expected polarizations of the top for top charged-Higgs production, which is
significantly different from that in the closely related process of $t W$
production in the Standard Model. We then show that an azimuthal asymmetry, constructed from the
decay lepton angular distribution in the laboratory frame, is a sensitive probe of top polarization 
and can be used to constrain parameters involved in top charged-Higgs production.

\vspace*{0.2in}
PACS:  12.60.Fr, 13.88.+e, 14.65.Ha, 14.80.Fd

\vfill

\section{Introduction}

The properties and interactions of the top quark, except for its mass, are not yet known in detail. With a mass close to the electroweak (EW) symmetry breaking scale and thus a large Yukawa coupling, the top quark is an excellent probe of whatever mechanism is responsible for EW symmetry breaking. In the Standard Model (SM), EW symmetry is broken through a single $SU(2)$ scalar doublet, i.e, through the Higgs mechanism. However, while the SM Higgs mechanism is the simplest way to break EW symmetry, there are reasons to consider an enlarged Higgs sector \cite{Gunion:1989we}. Models with two Higgs doublets can generate spontaneous $CP$ violation, address the strong $CP$ problem and generate additional sources of $CP$ violation needed for baryogenesis \cite{Branco:1999fs}. Moreover, the most popular paradigm for addressing the gauge hierarchy problem, supersymmetry (SUSY) contains two Higgs doublets in its simplest formulation \cite{Gunion:1989we, Djouadi:2005gj}. The spectrum of two Higgs doublet models (THDM) involves three neutral and two charged Higgs bosons. Different versions of the THDM also have different couplings of the scalars to fermions. Thus, even if scalar particles were to be  discovered at the Large Hadron Collider (LHC), it is necessary to probe in detail the precise couplings to these particles to establish the underlying model and pinpoint the exact mechanism of EW symmetry breaking. Charged Higgs particles exist even in
extensions of the SM which involve the introduction 
of a $SU(2)$ triplet of scalars, 
which are also interesting from the point of
view of obtaining a small Majorana mass for neutrinos in the type-II
see-saw mechanism \cite{type2seesaw}. It is possible to produce a single
top quark in association with a charged Higgs in such models. We study,
in this work, such a process in the context of a type II THDM or SUSY models, 
where the up type quarks couple to one of the Higgs doublets and down 
type quarks couple to the other Higgs doublet \cite{Gunion:1989we}.

The study of the top quark at the Tevatron has made use of the sample of
top-antitop pairs produced in large numbers. At the LHC, there would be
copious production of $t\bar t$ pairs, and one can think of the LHC as a top
factory. While pair production would be most useful for studying many
properties of the top quark, single-top production, which proceeds via the 
weak interaction, would be more suitable to study the weak sector. In
particular, measurement of the CKM matrix element $V_{tb}$ can be made
using single-top events. While a few
single-top events have been seen at the Tevatron, at the LHC a much
larger rate will be seen, and the single-top channel will be useful for a 
confirmation of the SM couplings for the top, and a precise measurement
of $V_{tb}$.

With a large mass of $\sim 172$ GeV, the top quark has an extremely short lifetime, calculated in the SM to be $\tau_t=1/\Gamma_t \sim 5 \times 10^{-25}$ s. This is an order of magnitude smaller than the hadronization time scale, which is roughly $1/\Lambda_{\rm{QCD}} \sim 3 \times 10^{-24}$ s. Thus, in contrast to lighter quarks, the top decays before it can form bound states with lighter quarks \cite{Bigi:1986jk}. 
As a result, the spin information of the {\it bare} top, which depends solely on its production process, is reflected in characteristic angular distributions of its decay products. Thus, the degree of polarization of an ensemble of top quarks can provide important information about the underlying physics in its production, apart from usual variables like cross sections, since any couplings of the top to new particles can alter its degree of polarization and the angular distributions of its decay products\footnote{For reviews on top quark physics and polarization see \cite{Bernreuther:2008ju,Beneke:2000hk,Wagner:2005jh}.}. In this paper, we investigate the effects on top polarization in the single production of the top in association with a charged Higgs of the type II THDM or the minimal supersymmetric standard model (MSSM). 

Single-top production in association with a charged Higgs can be used to
probe the size and nature of the $tbH$ coupling. Apart from the cross
section, the angular distribution of the top, and even the
polarization of the top would give additional information enabling the
determination of the $tbH$ coupling. 
Here we concentrate on the polarization of the top in the process, which
would be a measure of the extent of parity violation in the couplings.
It will be seen that polarization gives a handle on the combination 
$A_L^2 - A_R^2$
of the left-handed and right-handed couplings, 
 $A_L\equiv m_t\cot\beta$ and 
$A_R\equiv m_b\tan\beta$ of the
charged Higgs to the top where $\tan \beta$ is the ratio of the vacuum expectation values (vevs) of the Higgs doublets, in contrast to the combination $A_L^2 + A_R^2$
measured by the cross section or angular distribution. 

The most direct way to determine top polarization is by measuring the angular distribution of its decay products in its rest frame. However, at the LHC reconstructing the top rest frame will be difficult. In this paper, we show how the decay lepton angular distributions in the laboratory frame can be a useful probe of top polarization and the $tbH^-$ coupling. As will be explained in Section 2, the angular distribution of the charged lepton has a special property$-$it is independent of new physics in the $tbW$ decay vertex, to linear order in the anomalous couplings, and is thus a pure probe of new physics in top production alone. We show that the azimuthal distribution of the lepton is sensitive to top polarization and can be used to probe the coupling parameter $\tan \beta$ in the type II THDM. This approach has been recently
used to probe new physics in the case of top pair production in a model with an extra
heavy vector resonance $(Z')$ with chiral couplings \cite{Godbole:2010kr}. The effects of top polarization in $tW$ and $tH^-$ production have been studied previously in \cite{Beccaria:2004xk}, where the effects of 1-loop electroweak SUSY corrections have been considered; however, they do not consider top decay. Top polarization in different modes of single top production has also been studied in \cite{espriu}, where spin sensitive variables are used to analyze effective left and right handed couplings of the top coming from BSM physics.

This paper is organized as follows. In Section 2 we discuss top polarization and outline the spin density matrix formalism, needed to preserve spin coherence between top
production and decay. In Section 3, we derive expressions for polarized cross sections for $tH^-$ production and present results for the expected top polarization in this case. In Section 4, we construct an azimuthal asymmetry involving the charged lepton from top decay which is a probe of top polarization and a sensitive measure of $\tan \beta$.  Section 5 contains a summary.


\section{Top polarization and the spin density matrix}
Top spin can be determined by the angular distribution of its decay products. In the SM, the dominant decay mode is $t\to b W^+$, with a branching ratio (BR) of 0.998, with the $W^+$ subsequently decaying to $\ell^+ \nu_\ell$ (semileptonic decay, BR 1/9 for each lepton) or $u \bar{d}$,$c\bar{s}$ (hadronic decay, BR 2/3). The angular distribution of a decay product $f$ for a top quark ensemble has the form ( see for example  \cite{Bernreuther:2008ju}),
 \begin{equation}
 \frac{1}{\Gamma_f}\frac{\textrm{d}\Gamma_f}{\textrm{d} \cos \theta _f}=\frac{1}{2}(1+\kappa _f P_t \cos \theta _f).
 \label{topdecaywidth}
 \end{equation}
 Here $\theta_f$ is the angle between $f$ and the top spin vector in the top rest frame and
\begin{equation}
 P_t=\frac{N_\uparrow - N_\downarrow}{N_\uparrow + N_\downarrow}, 
 \label{ptdef}
 \end{equation}
 is the degree of polarization of the top quark ensemble where $N_\uparrow$ and $N_\downarrow$ refer to the number of positive and negative helicity tops respectively. $\Gamma_f$ is the partial decay width and $\kappa_f$ is the spin analyzing power of $f$. Obviously, a larger $\kappa_f$ makes $f$ a more sensitive probe of the top spin. The charged lepton and $d$ quark are the best spin analyzers with $\kappa_{\ell^+}=\kappa_{\bar{d}}=1$, while $\kappa_{\nu_\ell}=\kappa_{u}=-0.30$ and  $\kappa_{b}=-\kappa_{W^+}=-0.39$, at tree level \cite{Bernreuther:2008ju}. Thus the $\ell^+$ or $d$ have the largest probability of being emitted in the direction of the top spin and the least probability in the direction opposite to the spin. Since at the LHC, leptons can be measured with high precision, we focus on leptonic decays of the top.
 
 For hadronic $t\bar{t}$ production, spin correlations between the decay leptons from the $t$ and $\bar{t}$ have been extensively studied in the SM and for BSM scenarios \cite{Bernreuther:2008ju,Beneke:2000hk,spincorr}. These spin correlations measure the asymmetry between the production of like and unlike helicity pairs of $t\bar{t}$ which can probe new physics in top pair production. However, this requires the reconstruction of the $t$ and $\bar{t}$ rest frames, which is difficult at the LHC. Here we investigate top polarization in the lab. frame, which would be more directly and easily measurable without having to construct the top rest
frame.

Let us consider a generic process of top charged-Higgs production and subsequent 
semileptonic decay of $t$ and inclusive decay of $H^-$, $A B \to t H^- \to b \ell^+ \nu_\ell X$. Since $\Gamma_t/m_t \sim 0.008$, we 
can use the narrow width approximation (NWA) to write the cross section as 
a product of the $2\to 2$
production cross section times the decay 
width of the top. However, in probing top polarization using angular 
distributions of the decay lepton, it is necessary to keep the top 
spin information in its decay arising from its production, thus requiring the spin 
density matrix formalism. As in \cite{Godbole:2006tq}, the amplitude squared can be factored into production and decay parts using the NWA as 
\begin{eqnarray}
\overline{|{\cal M}|^2} = \frac{\pi \delta(p_t^2-m_t^2)}{\Gamma_t m_t}
\sum_{\lambda,\lambda'} \rho(\lambda,\lambda')\Gamma(\lambda,\lambda'),
\end{eqnarray}
where $\rho(\lambda,\lambda')$ and $\Gamma(\lambda,\lambda')$ are the $2 \times 2$ top production and decay spin density matrices and $\lambda,\lambda' =\pm 1$ denote the sign of the top helicity. After phase space integration of $\rho(\lambda,\lambda')$ we get the resulting polarization density matrix $\sigma(\lambda,\lambda^\prime)$. The (1,1) and (2,2) diagonal elements of $\sigma(\lambda,\lambda^\prime)$ are the cross sections  for the production of positive and negative helicity tops and $\sigma_{\rm{tot}}=\sigma(+,+)+\sigma(-,-)$ is the total cross section. We define the degree of {\it longitudinal} polarization $P_t$ as
\begin{equation}
P_t=\frac{\sigma(+,+)-\sigma(-,-)}{\sigma(+,+)+\sigma(-,-)}.
\label{eta3def}
\end{equation}
The off-diagonal elements of $\sigma(\lambda,\lambda^\prime)$ are the production rates of the top with {\it transverse} polarization. The top decay density matrix $\Gamma(\lambda,\lambda')$ for the process $t\to b W^{+}\to b \ell^{+} \nu_\ell$ can be 
written in a Lorentz invariant form as
\beq
\Gamma(\pm,\pm)=2 g^4 \ |\Delta(p_W^2)|^2 (p_b \cdot p_\nu)
 \left[(p_\ell \cdot p_t) \mp m_t (p_\ell \cdot n_3)\right],
\label{tdecaydia}
\eeq
for the diagonal elements and
\beq
\Gamma(\mp,\pm)=-2 g^4 \ |\Delta(p_W^2)|^2 
\,m_t \,\,(p_b \cdot p_\nu) \,\, p_\ell \cdot (n_1 \mp i n_2),
\label{tdecayoffdia}
\eeq
for the off-diagonal ones. Here $\Delta(p_W^2)$ is the $W$ boson propagator and $n^{\mu}_{i}$'s ($i=1,2,3$) are the 
spin 4-vectors for the top with 4-momentum $p_t$, with the properties 
$n_i \cdot n_j =-\delta_{ij}$ and $n_i \cdot p_t =0$. 
For decay in the rest frame they take the standard form $n^{\mu}_{i}=(0, 
\delta_{i}^{k})$. 

 Using the NWA the differential cross section for top production and decay, with inclusive decay of $H^-$ can be written as 
\begin{eqnarray}
d\sigma&=&\frac{1}{32 \ \Gamma_t m_t} \ \frac{1}
{(2\pi)^4} \left[ \sum_{\lambda,\lambda'} d\sigma(\lambda,\lambda') \
\times \left(\frac{\Gamma(\lambda,\lambda')}{p_t \cdot p_\ell}\right) \right] E_\ell \ |\Delta(p_W^2)|^2 \
d\cos\theta_t \ d\cos\theta_\ell \ d\phi_\ell\nonumber \\
&\times& dE_\ell \ dp_W^2,
\label{dsigell}
\end{eqnarray}
where the lepton integration variables are in the lab frame and $b$ quark energy integral is replaced by an integral over the invariant mass $p_W^2$ of the $W$ boson. $d\sigma(\lambda,\lambda')$ is the differential cross section for the $2\to2$ process of top charged Higgs production with indicated spin indices of the top. As shown in \cite{Godbole:2006tq}, by measuring the angular distributions of the decay lepton in the top rest frame (which requires reconstructing the top rest frame) analytic expressions for the longitudinal and transverse components of the top polarization can be obtained by a suitable combination of lepton polar and azimuthal asymmetries. However, as pointed out in the introduction, it would be useful and interesting to devise variables for the lepton in the laboratory frame, which are easily measured and are sensitive to top polarization.

An important point is the possible appearance of new physics in the $tbW$ decay vertex, apart from that in top production, leading to changed decay width and distributions for the $W^+$ and $l^+$. The $tbW$ vertex can be written in model-independent form as
\begin{equation}
 \Gamma^\mu =\frac{-ig}{\sqrt{2}}\left[\gamma^{\mu}(f_{1L} P_{L}+f_{1R}P_{R})-\frac{i \sigma^{\mu \nu}}{m_W}(p_t -p_b)_{\nu}(f_{2L}P_{L}+f_{2R}P_{R})\right], \label{anomaloustbW}
\end{equation}
where for the SM $f_{1L}=1$ and the anomalous couplings $f_{1R}=f_{2L}=f_{2R}=0$. 
The simultaneous presence of new physics in top production and decay can complicate the analysis making it difficult to isolate new couplings of the top. However, it has been proven that the energy averaged {\it angular distributions} of charged leptons or $d$ quarks from top decay are {\it not affected by the anomalous $tbW$ vertex}. This has been shown very generally for a $2\to n$ process and assumes the narrow width approximation (NWA) for the top and neglects terms quadratic in the anomalous couplings in (\ref{anomaloustbW}) assuming new physics couplings to be small (for details see \cite{Godbole:2006tq} and references therein). This implies that charged lepton angular distributions in the lab frame are more accurate probes of top polarization, and thus to new physics in top production alone. In contrast, the energy distributions of the $l^+$ or the angular distributions of the $b$ and $W$ are ``contaminated'' by the anomalous $tbW$ vertex. In section 4 we will construct an observable using the azimuthal distribution of the charged lepton which is sensitive to the top polarization and can be measured with a large significance at the LHC.

\section{Top polarization in the two Higgs doublet model}
\begin{figure}[htb]
\begin{center}
\begin{picture}(800,130)(0,0)

\Gluon(100,100)(140,60){5}{5}
\ArrowLine(100,20)(140,60)
\Vertex(140,60){2}
\ArrowLine(140,60)(190,60)
\Vertex(190,60){2}
\DashLine(190,60)(220,100){5}
\ArrowLine(190,60)(220,20)
\put(75,85){$g (p_1)$}
\put(75,35){$b (p_2)$}
\put(225,85){$H^- (p_4)$}
\put(225,35){$t (p_3,\lambda_t)$}
\put(160, 45){$b$}
\put(150,00){$(a)$}
\Gluon(310,100)(350,80){5}{4}
\ArrowLine(350,80)(390,100)
\Vertex(350,80){2}
\ArrowLine(350,50)(350,80)
\Vertex(350,50){2}
\DashLine(350,50)(390,20){5}
\ArrowLine(310,20)(350,50)
\put(285,85){$g (p_1)$}
\put(285,35){$b (p_2)$}
\put(390,35){$H^- (p_4)$}
\put(390,85){$t (p_3,\lambda_t)$}
\put(360,60){$t$}
\put(350,00){$(b)$}

\end{picture}
\caption{\sl Feynman diagrams contributing to the 
top charged-Higgs production at the LHC.}\label{feyngraph}
\end{center}
\end{figure}
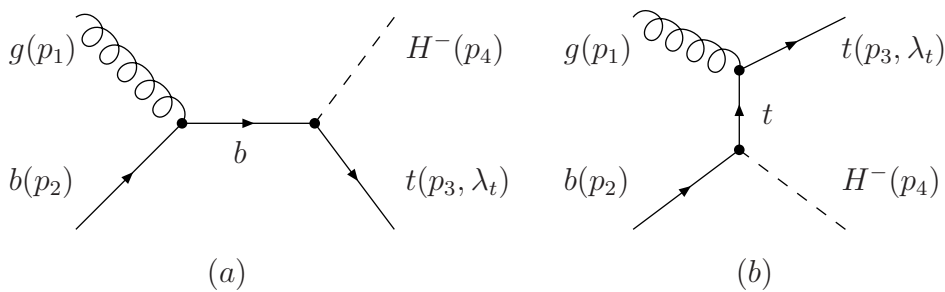

\begin{figure}[htb]
\begin{center}
\includegraphics[angle=270,width=3.2in]{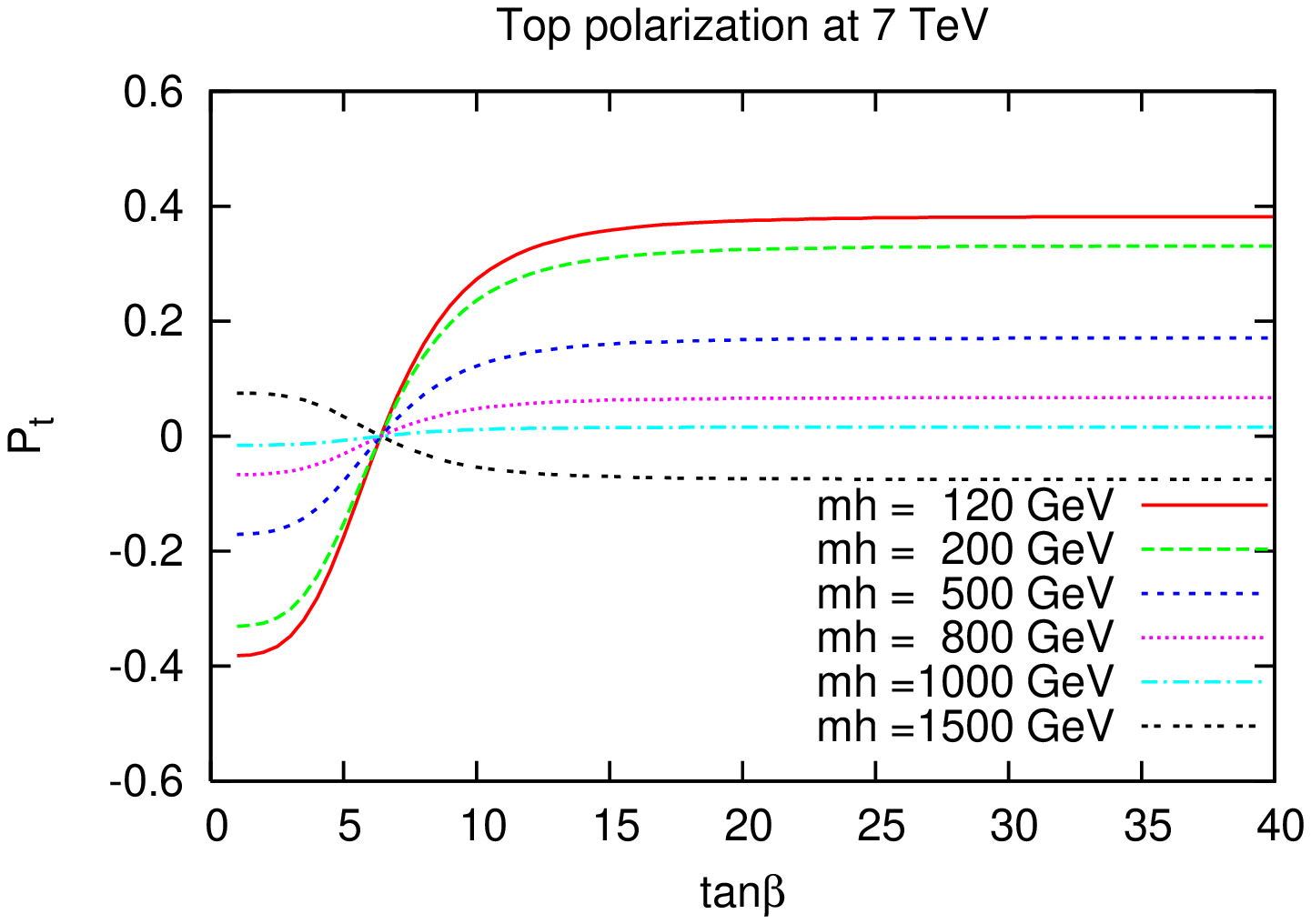} 
\includegraphics[angle=270,width=3.2in]{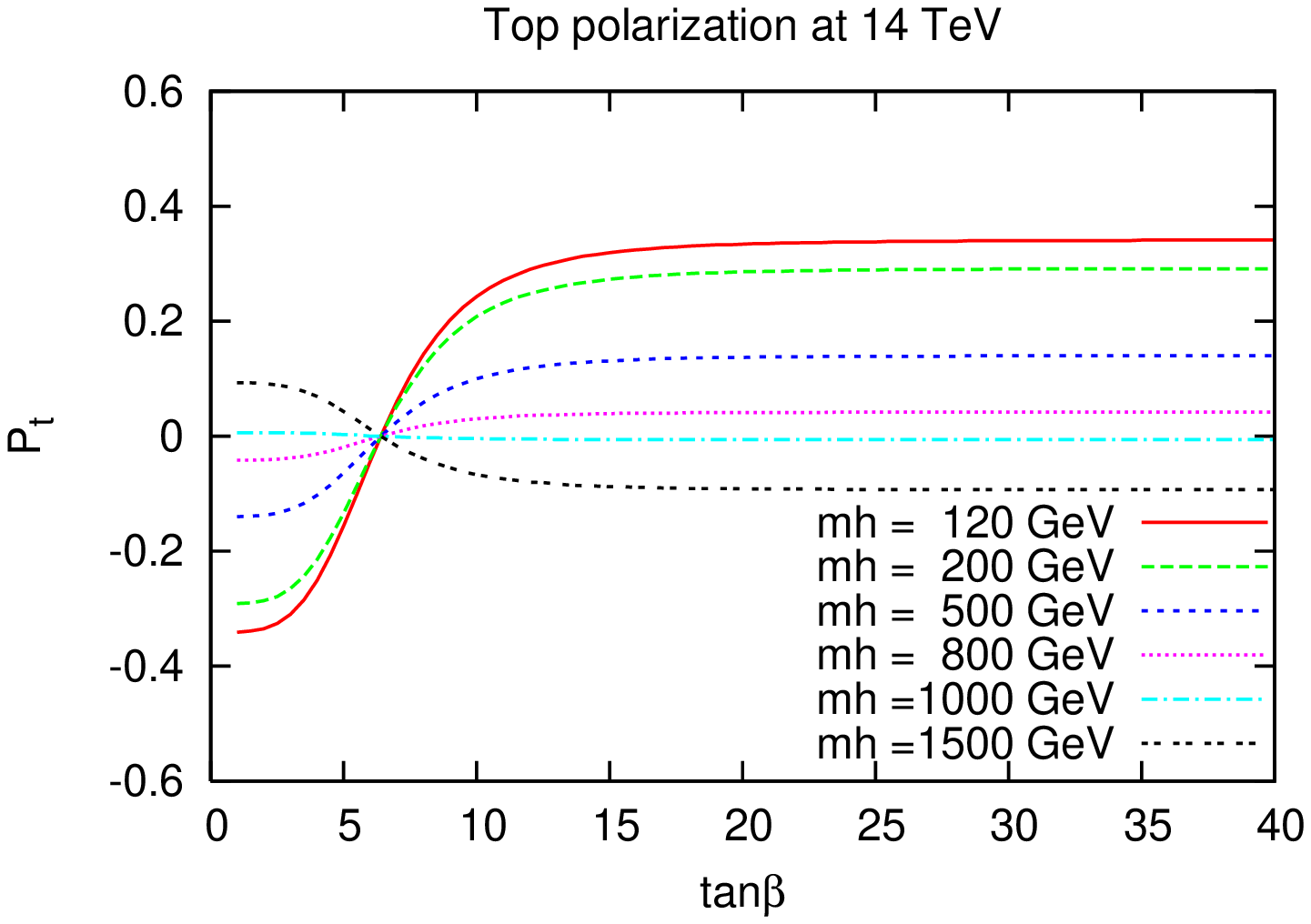} 
\caption{\sl The polarization asymmetries for top charged-Higgs production at LHC
for two different cm energies, 7 TeV (left) and 14 TeV (right), as a function of $\tan\beta$ for various charged Higgs masses.} 
\label{poltanb}
\end{center}
\end{figure}

We consider the process of single top production in association with a charged Higgs in the type II THDM or the Minimal Supersymmetric Standard Model (MSSM). For our purposes, the model is completely characterized by two parameters, the mass of the charged Higgs $M_{H^-}$ and the ratio of the vacuum expectation values (vevs) of the Higgs doublets $\tan \beta$. At the parton level, single top production proceeds via
\begin{equation}
 g(p_1)\, b(p_2)\rightarrow t(p_3,\lambda_t) H^{-}(p_4),
\end{equation}
where $\lambda_t=\pm 1$ is the sign of the helicity of the top. The tree level $s$ and $t$ channel diagrams contributing to the above process are shown in Fig.~\ref{feyngraph}.
%
%
As mentioned in the previous section, a study of top polarization using angular distributions of the top decay products requires computing the spin density matrix for top production and decay. We have obtained simple analytic expressions for the top production density matrix. In the type II THDM the $tbH^-$ coupling is 
\begin{equation}
g_{tbH^-}=\frac{g}{\sqrt{2} m_W} (m_t \cot \beta P_L + m_b \tan \beta P_R),
\label{tbhcoupling} 
\end{equation}
 where $g$ is the $SU(2)$ gauge coupling and $P_L$ and $P_R$ are the left and right handed projection operators respectively, $P_{L,R}=(1 \mp \gamma^5)/2$. One can immediately see that at $\tan \beta=\sqrt{m_t/m_b}$ , the pseudoscalar part of the coupling, which is proportional to $\gamma^5$, vanishes and the coupling (\ref{tbhcoupling}) is purely scalar.  Since polarization is parity violating we expect that the polarized cross section (\ref{eta3def}) should vanish for this value of $\tan \beta$ and we indeed find this to be the case, as will be shown later in Fig. \ref{poltanb}. 

Denoting the energy, momentum and scattering angle of the top in the parton center-of-mass (cm) frame by $E_t$, $p_t$ and $\theta_t$ respectively and the parton level Mandelstam variable by $\hat{s}$, the diagonal elements are given by
\begin{align}
 \rho(+,+)=F_1 \, m_t^2 \cot^2 \beta + F_2 \, m_b^2 \tan^2 \beta \\
 \rho(-,-)=F_2 \, m_t^2 \cot^2 \beta + F_1 \, m_b^2 \tan^2 \beta,
\label{diagonalrho} 
\end{align}
where $F_1$ and $F_2$ are defined by
\begin{align}
 F_1&= \left(\frac{g g_s}{2m_W}\right)^2 \frac{1}{6 \sqrt{\hat{s}}(E_t -p_t \cos \theta_t)^2}\left\lbrace p_t^2 (E_t+p_t)\sin^2 \theta_t \cos^2 \frac{\theta_t}{2}+ \left[4 E_t (E_t+p_t)(E_t -\sqrt{\hat{s}})\right. \right. \nonumber \\
&\hskip -1pt \left. \left. +\, 2 m_t^2 \sqrt{\hat{s}}+ (\hat{s}(E_t+p_t)+m_t^2(E_t-p_t)-4 m_t^2 E_t)\right] \sin ^2 \frac{\theta_t}{2} \right\rbrace  \\
F_2&= \left(\frac{g g_s}{2m_W}\right)^2 \frac{1}{6 \sqrt{\hat{s}}(E_t -p_t \cos \theta_t)^2}\left\lbrace p_t^2 (E_t-p_t)\sin^2 \theta_t \sin^2 \frac{\theta_t}{2}+ \left[4 E_t (E_t-p_t)(E_t -\sqrt{\hat{s}})\right. \right. \nonumber \\
&\hskip -1pt \left. \left. +\, 2 m_t^2 \sqrt{\hat{s}}+ (\hat{s}(E_t-p_t)+m_t^2(E_t+p_t)-4 m_t^2 E_t)\right] \cos ^2 \frac{\theta_t}{2} \right\rbrace.
\end{align}

The off-diagonal elements are
\begin{align}
 \rho(+,-)=\rho(-,+)&=-\left(\frac{g g_s}{2m_W}\right)^2 \frac{1}{6 \sqrt{\hat{s}}(E_t -p_t \cos \theta_t)^2}(m_t^2 \cot^2 \beta -m_b^2 \tan^2 \beta) \nonumber \\
& \hskip 0.6cm \times m_t \sin \theta_t (2 E_t \sqrt{\hat{s}} -m_t^2 -\hat{s} +p_t^2 \sin^2 \theta_t).
\label{nondiagonalrho}
\end{align}
 
\begin{figure}[htb]
\begin{center}
\includegraphics[angle=270,width=3.2in]{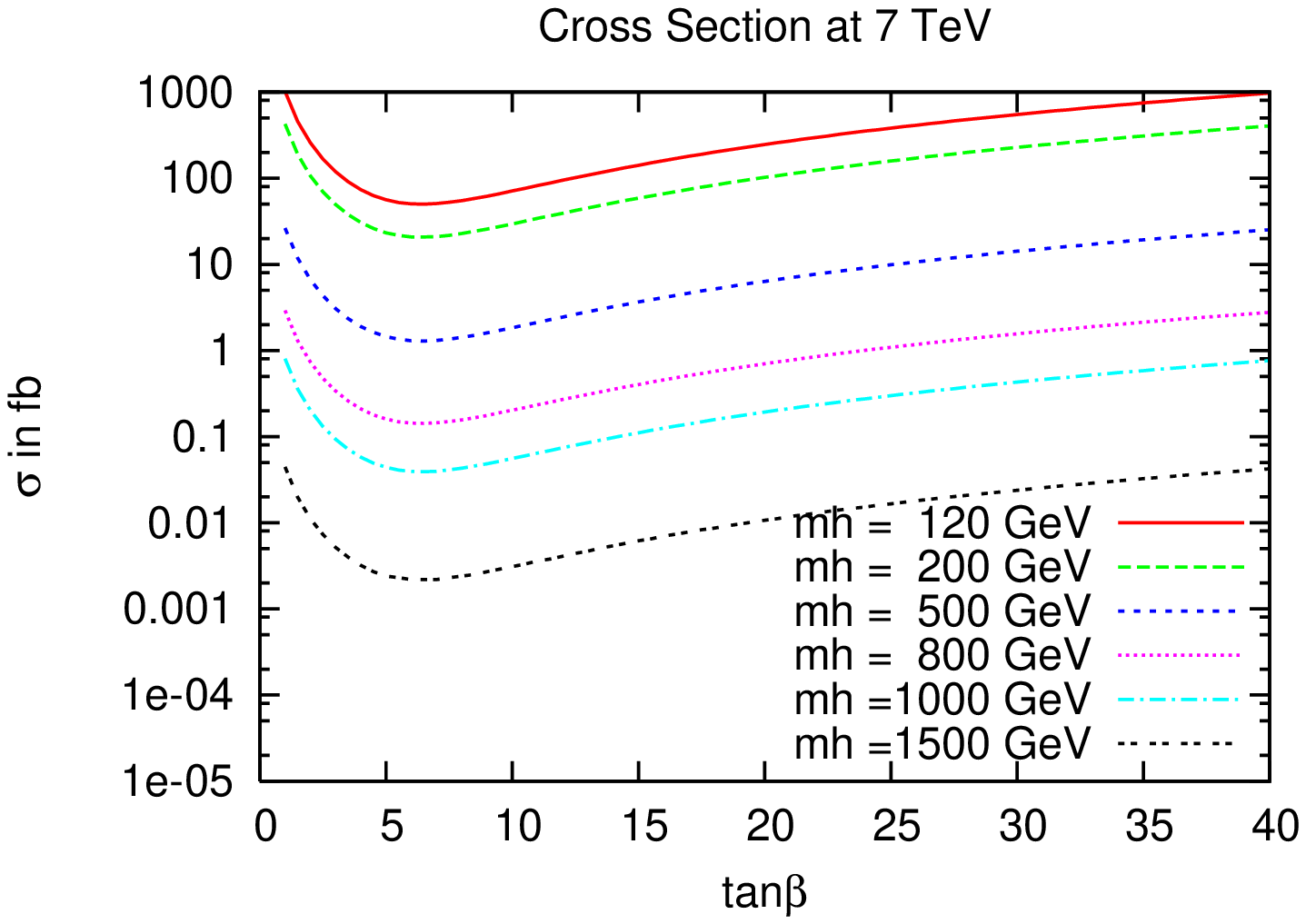} 
\includegraphics[angle=270,width=3.2in]{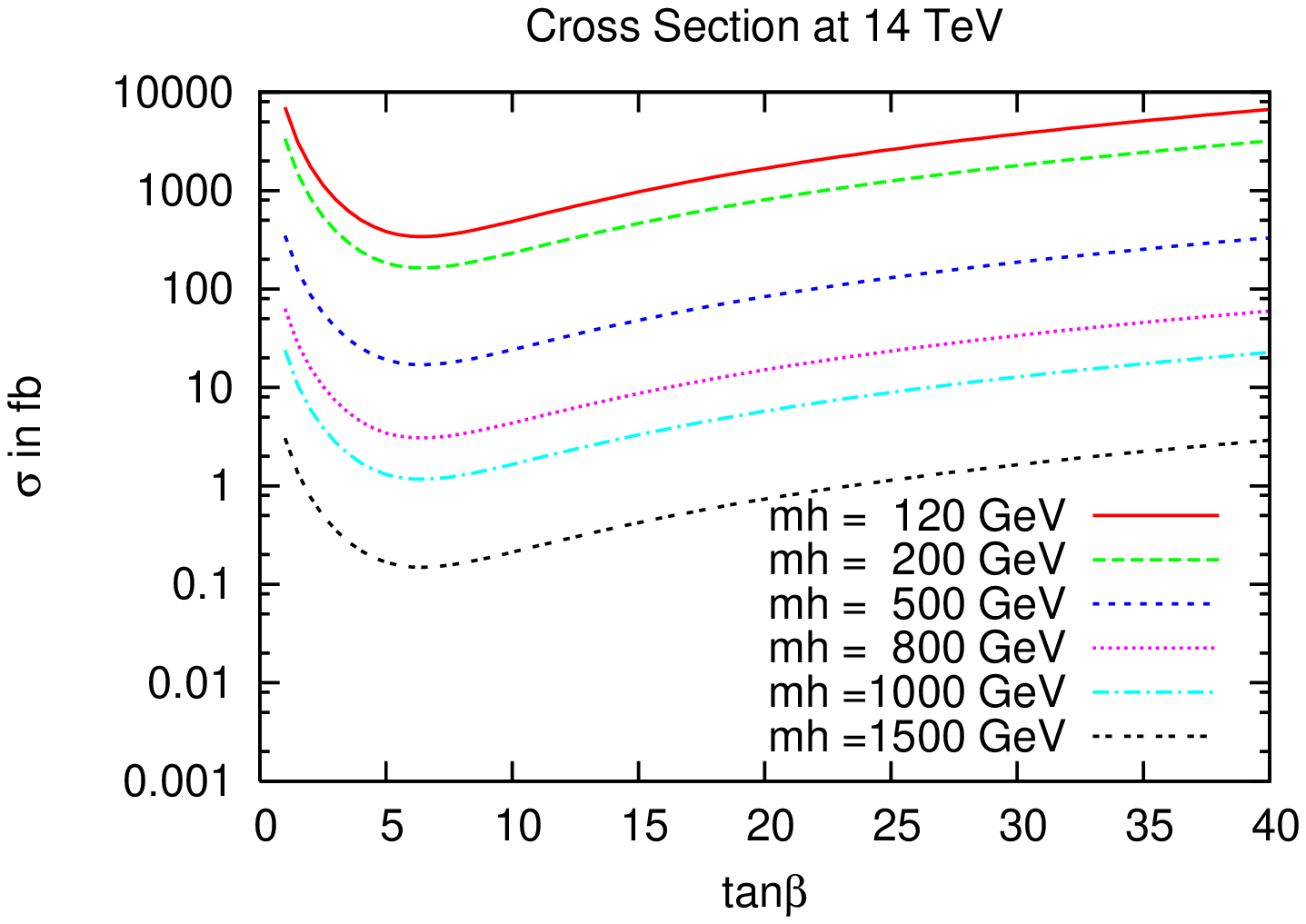} 
\caption{\sl The cross section for top charged-Higgs production at LHC for 
two different cm energies, 7 TeV (left) and 14 TeV (right), as a function of $\tan \beta$ for various charged Higgs masses.} 
\label{cs}
\end{center}
\end{figure}

In deriving the above expressions we have neglected the kinematic effects of the $b$ quark mass but kept factors of $m_b$ occurring in the $tbH^-$ coupling (\ref{tbhcoupling}). Analytic expressions for the helicity amplitudes for associated $tH^-$ production can be found in \cite{Beccaria:2004xk}, where a similar convention for retaining factors of $m_b$ is used; our density matrix elements (\ref{diagonalrho}) and (\ref{nondiagonalrho}), obtained by an independent method, agree with those obtained using the helicity amplitudes of \cite{Beccaria:2004xk}. A plot of the cross section as a function of the coupling $\tan \beta$ is shown in Fig. \ref{cs} for various values of charged Higgs masses. We show the cross section for two different center of mass energies of 7 TeV and 14 TeV for which the LHC is planned to operate and have used the leading order parton density function (PDF) sets of CTEQ6L1 \cite{cteq6l}. We see that the cross sections have a similar profile for various $M_{H^-}$ values and fall sharply for larger $M_{H^-}$. The cross sections are proportional to $(m_t^2 \cot^2 \beta +m_b^2 \tan ^2 \beta)$, which is minimized for $\tan \beta=\sqrt{\frac{m_t}{m_b}}\simeq 6.41$, independent of the center-of-mass energy and the value of $M_{H^-}$. This can indeed be seen from Fig. \ref{cs}. Here we have taken the top mass to be 172.6 GeV and have evaluated the PDF's at the same scale.

The $tbH^-$ vertex has a scalar-pseudoscalar $(A+B \gamma^5)$ chiral structure which is different from vector-axial vector coupling of the $tbW$ and $t\bar{t}Z^0$ vertices. One thus expects a very different longitudinal polarization asymmetry given by Eqn. (\ref{eta3def}) for top charged-Higgs production compared to $t\bar{t}$ production, and for the closely related process of associated $tW$ production in the SM proceeding via $gb \to t W$. For SM $tW$ production we find the longitudinal polarization to be $P_t \simeq -0.25$; for $t \bar{t}$ production it is $\mathcal{O}(-10^{-4})$. The very small value of $P_t$ for top pair production in the SM is because the dominant contribution for both $gg \to t \bar{t}$ and $q\bar q\to t \bar t$ comes from chirality conserving $s$-channel gluon exchange processes, resulting in the production of largely unpolarized tops. These values of $P_t$ have also been calculated in \cite{Arai:2010ci}, where top polarization effects for top-slepton production in $R$-parity violating SUSY was considered. We show the polarization asymmetry for $tH^-$ production in Fig. \ref{poltanb} as a function of $\tan\beta$ for both $\sqrt{s}=7$ and 14 TeV.
In contrast to the related case of top-slepton production considered in \cite{Arai:2010ci} where $P_t$ was found to be independent of the $R$-parity violating SUSY $tb\tilde{l}$ coupling, here $P_t$ does have an interesting dependence on $\tan\beta$. As mentioned previously, we notice the interesting feature that the polarization vanishes at $\tan\beta=\sqrt{\frac{m_t}{m_b}}$ for all $M_{H^-}$ and $\hat{s}$, as expected from the vanishing of the chiral part of the coupling (\ref{tbhcoupling}) at this $\tan\beta$ value, the same value for which the cross sections are minimized. The curves change sign at this point and saturate rapidly for larger $\tan\beta$ values.  

\begin{figure}
\begin{center}
\includegraphics[angle=270,width=3.2in]{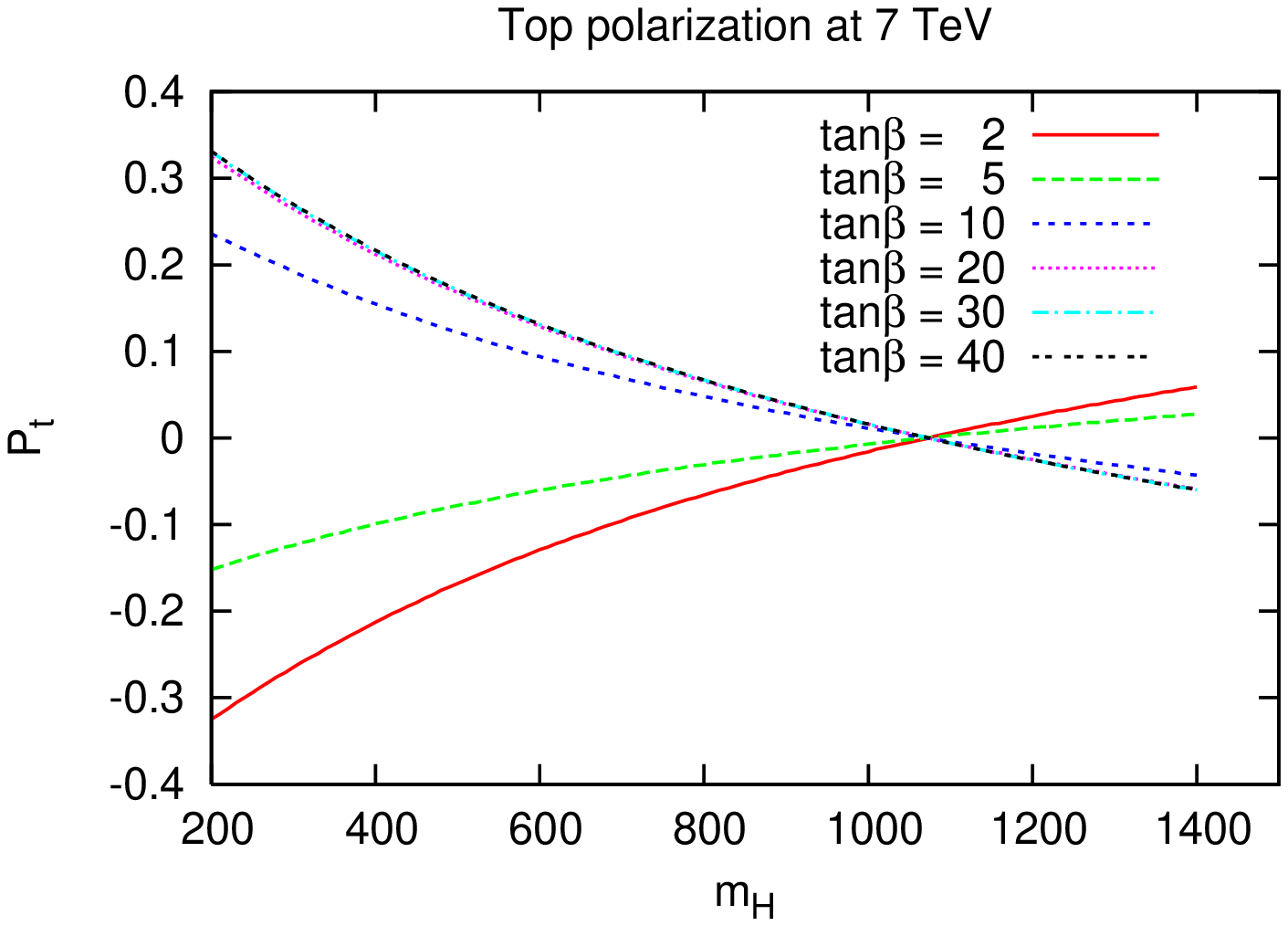}
 \includegraphics[angle=270,width=3.2in]{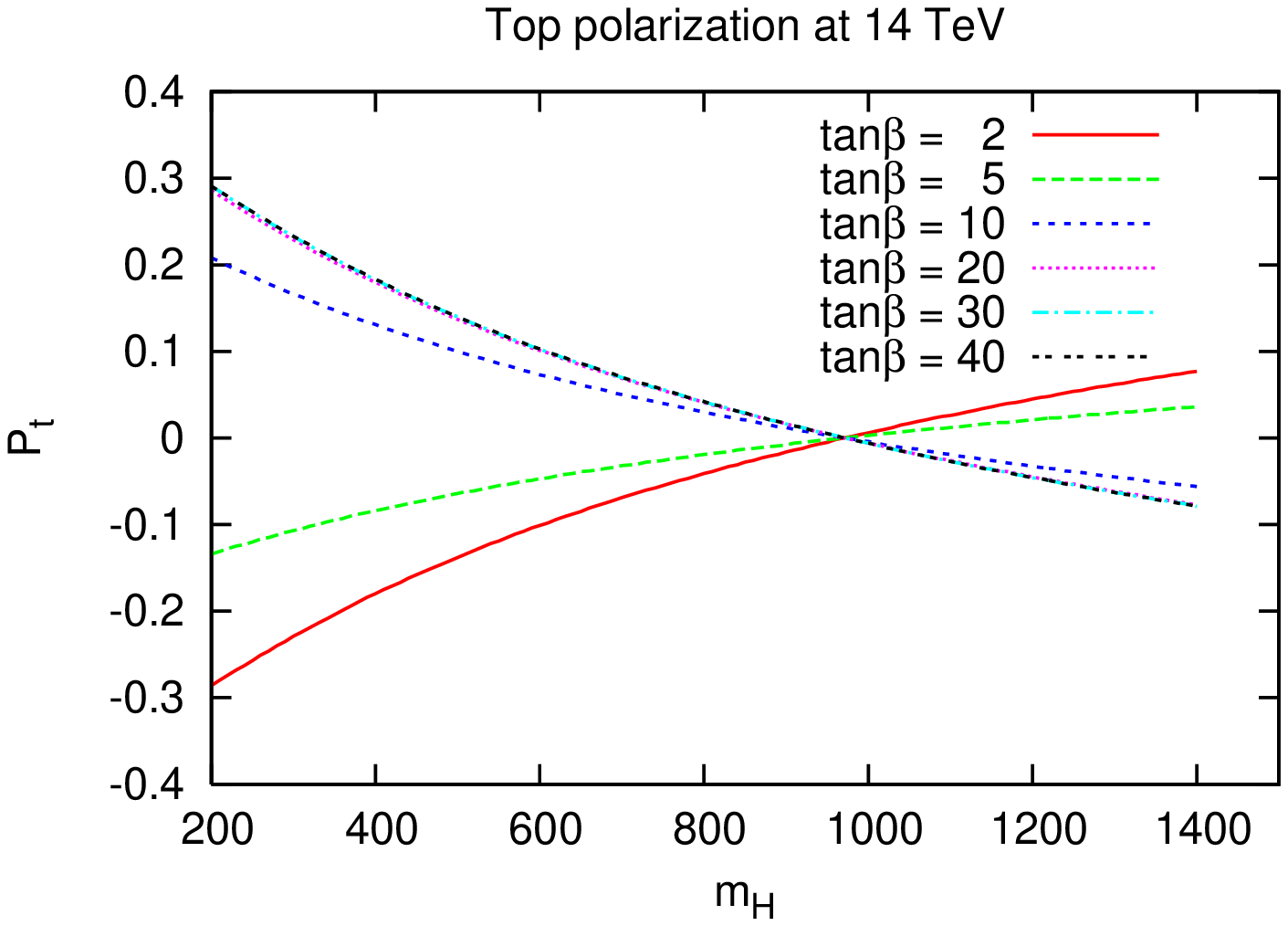} 
\caption{\sl The polarization asymmetry for top charged-Higgs production at LHC
for a cm energy of 7 TeV (left) and 14 TeV (right), as a function of $M_{H^-}$ for various $\tan\beta$ values.} 
\label{polmh}
\end{center}
\end{figure}

A plot of $P_t$ vs the charged Higgs mass for various values of $\tan\beta$ is shown in Fig. \ref{polmh}, for $\sqrt{s}=7$ and 14 TeV.
We notice that the polarization asymmetry vanishes for a charged Higgs mass close to 1100 GeV for $\sqrt{s}=7$ TeV and around 1000 GeV for the 14 TeV case, for all $\tan\beta$, and changes sign as $M_{H^-}$ is increased. This can be understood as follows. In the expression for the polarization asymmetry $P_t \propto \rho(+,+)-\rho(-,-)$, the angular integrals can be done analytically. Since the parton distributions of the gluon and $b$ quark peak at low $x$, the remaining PDF integrals over the momentum fractions of the gluon and $b$ are dominated at low $x$, i.e, at the threshold for top charged-Higgs production. One can show that the expressions for $P_t$, expanded in powers of the top momentum $p_t$ (i.e, evaluated close to $\hat{s}=(m_t +M_{H^-})^2$), vanishes for $M_{H^-}=6 m_t \simeq 1035.6$ GeV at leading order in $p_t$, for all $\tan \beta$, in reasonable agreement with Fig. \ref{polmh}. Of course, one cannot get an exact analytic expression for $M_{H^-}$ when $P_t$ vanishes without doing the numerical integrals over the gluon and $b$ quark PDF's. Still, the above argument, which is independent of the center-of-mass energy of the colliding protons, is useful for understanding why the polarization vanishes close to $M_{H^-}\simeq 1000$ GeV for both $\sqrt{s}=7$ and 14 TeV.  

The important point to note is that the magnitude and sign of these asymmetries are sensitively dependent on $M_{H^-}$ and $\tan \beta$ values and are significantly different from the case of $tW$ and $t\bar{t}$ production, because of the different chiral structure of the $tbW$ vertex.

\section{Azimuthal distributions of decay leptons}
As mentioned in previous sections, the top quark decays rapidly and its properties have to be deduced from its decay products. The top polarization can be determined by the angular distribution of its decay products using Eqn. (\ref{topdecaywidth}). The lab frame polar distribution of the lepton is independent of the anomalous $tbW$ decay vertex. However, we find that it is not sensitive to model parameters and is largely indistinguishable from the $tW$ case in the SM. 

As shown in \cite{Godbole:2006tq} and references therein, the azimuthal angle of the decay lepton in the lab frame is sensitive to the top polarization and independent of possible new physics in the $tbW$ decay vertex and is thus a convenient probe. The lepton azimuthal angle $\phi_\ell$ is defined with respect to the top production plane chosen as the $x-z$ plane, with the beam direction as the $z$ axis and the convention that the $x$ component of the top momentum is positive. Since at the LHC, one cannot uniquely define a positive direction of $z$ axis, the lepton azimuthal distribution is identical for $\phi_l$ and $2 \pi -\phi_l$ and is symmetric around $\phi_l=\pi$. 

The $\phi_\ell$ distributions for pure, i.e, $100\%$, positively or negatively polarized top quark ensemble is obtained by using only the $(+,+)$ or $(-,-)$ density matrix elements respectively in Eqn. (\ref{dsigell}). This is, of course, expected to be different from that for an ensemble with a partial degree of polarization $P_t$. In computing the $\phi_\ell$ distributions we have taken into account the full spin coherence effects of the top encoded in the diagonal and off-diagonal elements of the production and decay spin density matrices. 

\begin{figure}
\begin{center}
\includegraphics[angle=270,width=3.2in]{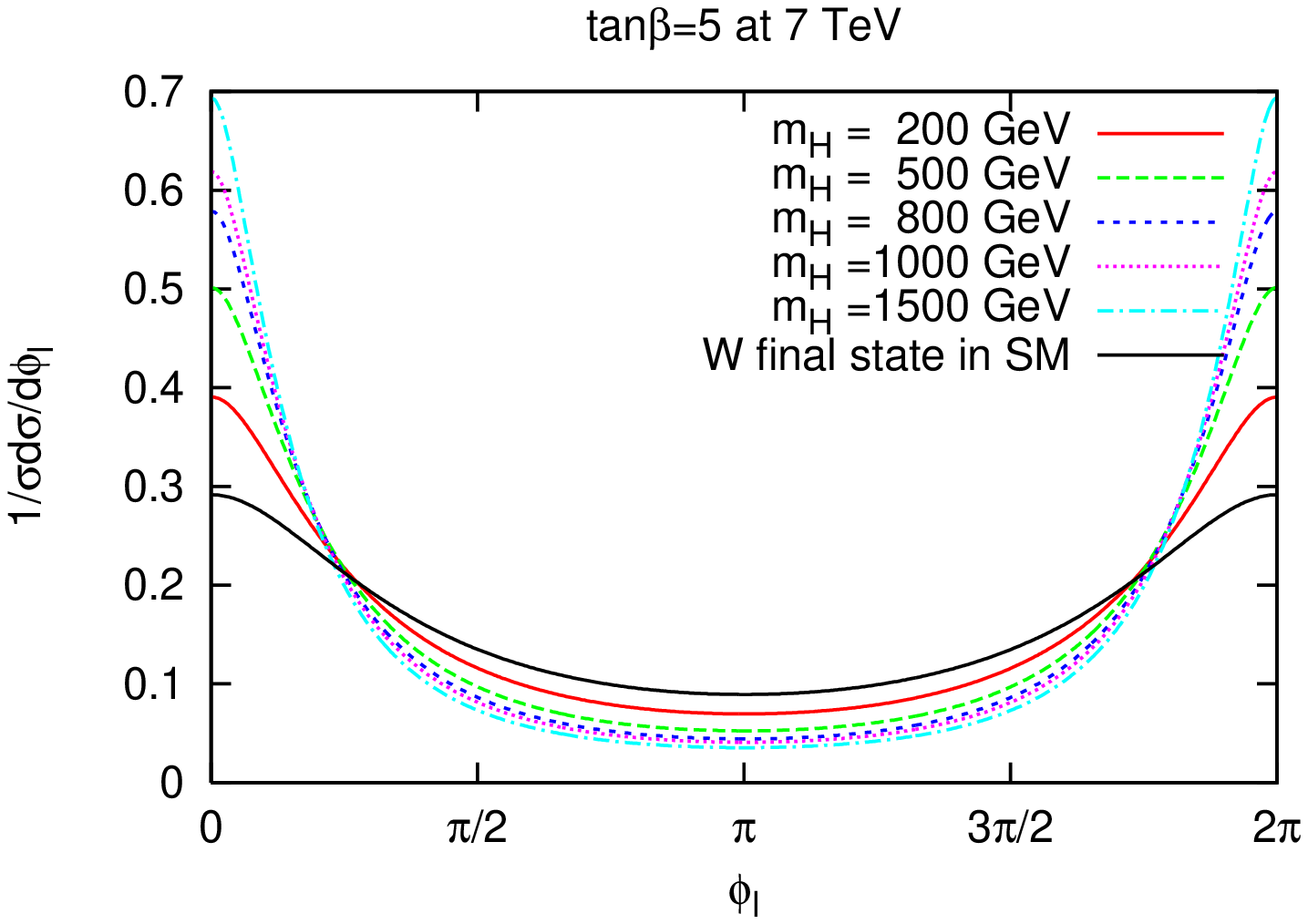}
 \includegraphics[angle=270,width=3.2in]{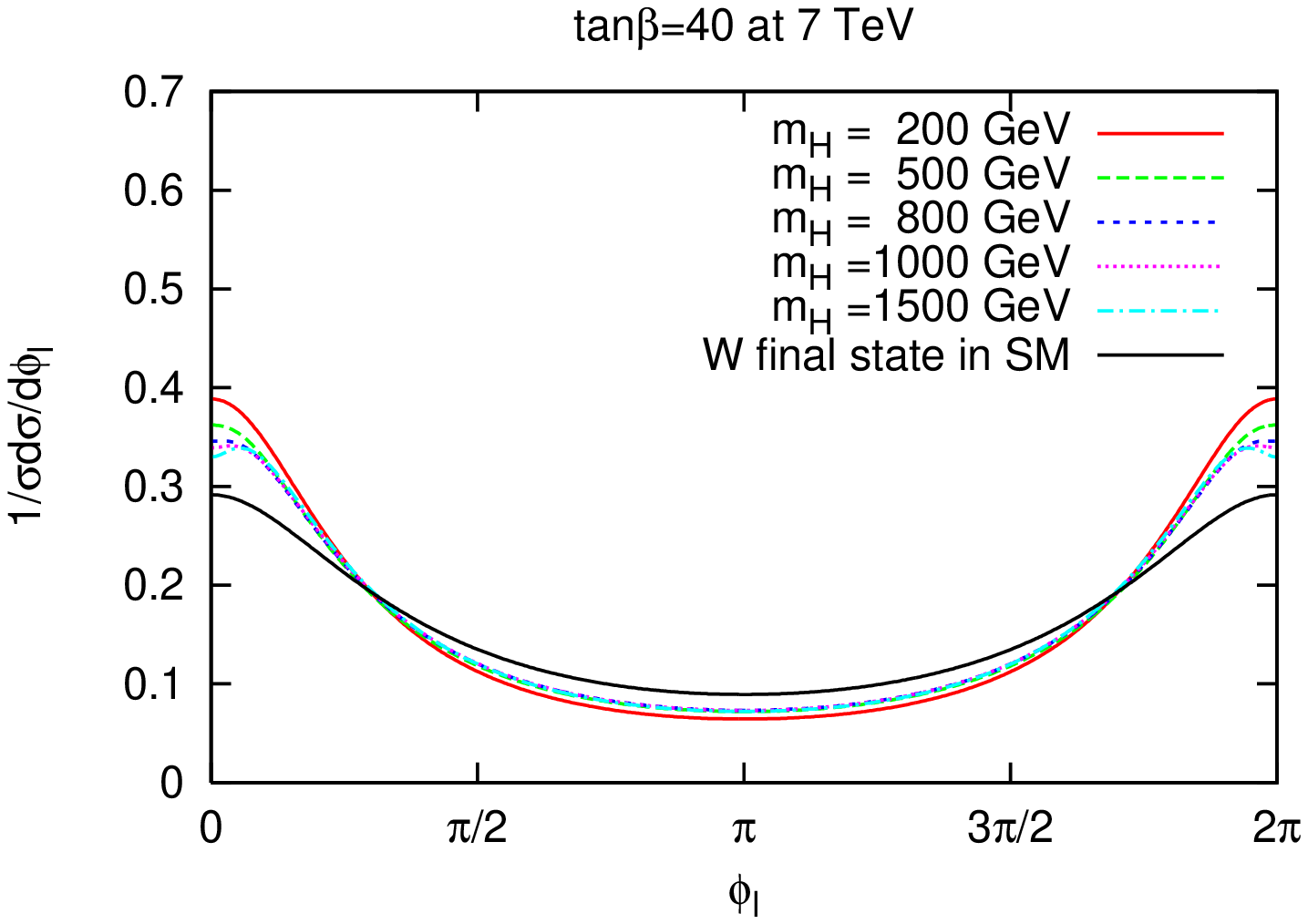} 
\caption{\sl The normalized lepton azimuthal distribution for $\tan \beta=5$ (left) and $\tan\beta=40$ (right) for various charged Higgs masses at a cm energy of 7 TeV.} 
\label{phidis7TeV}
\end{center}
\end{figure}

\begin{figure}
\begin{center}
\includegraphics[angle=270,width=3.2in]{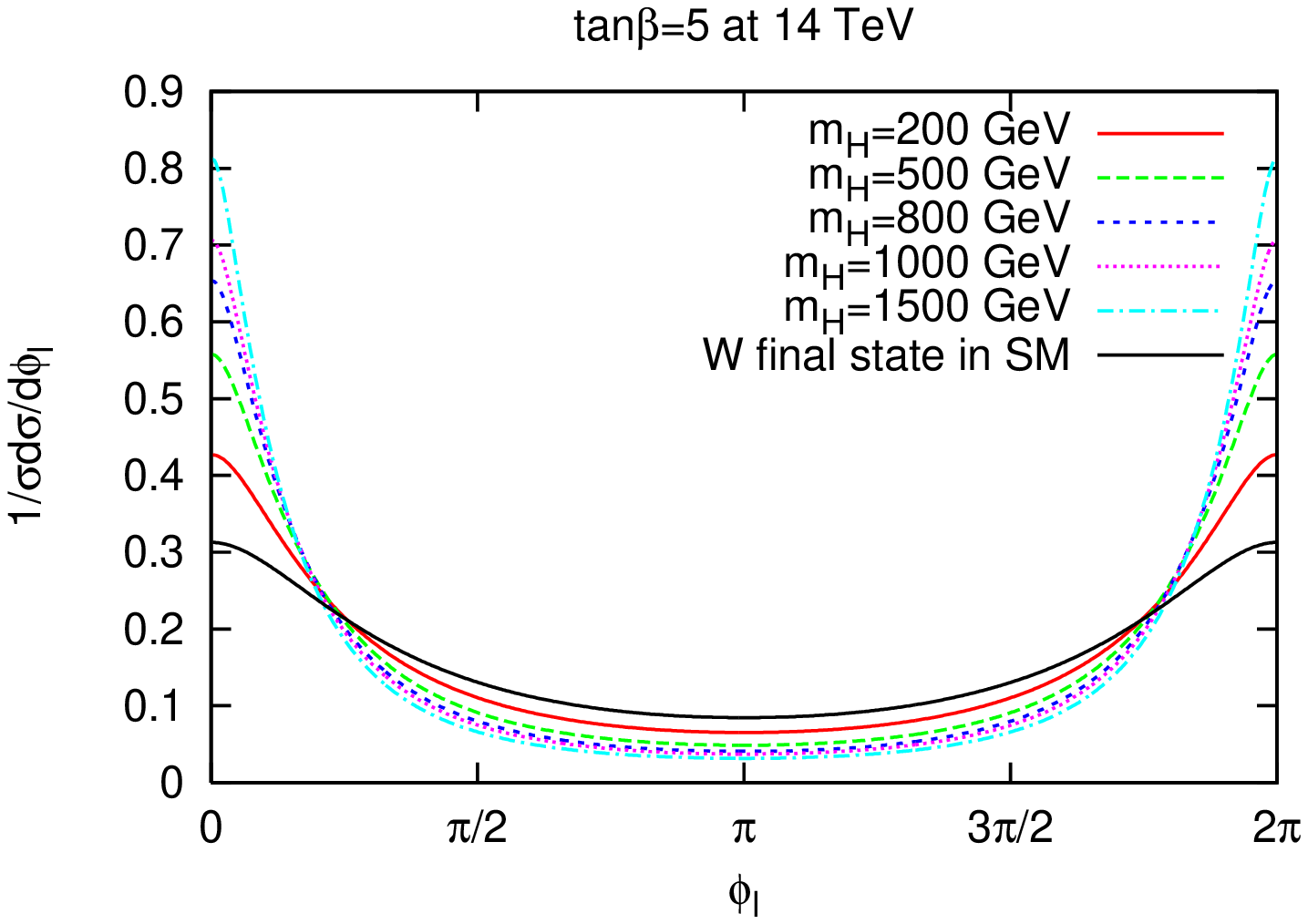}
 \includegraphics[angle=270,width=3.2in]{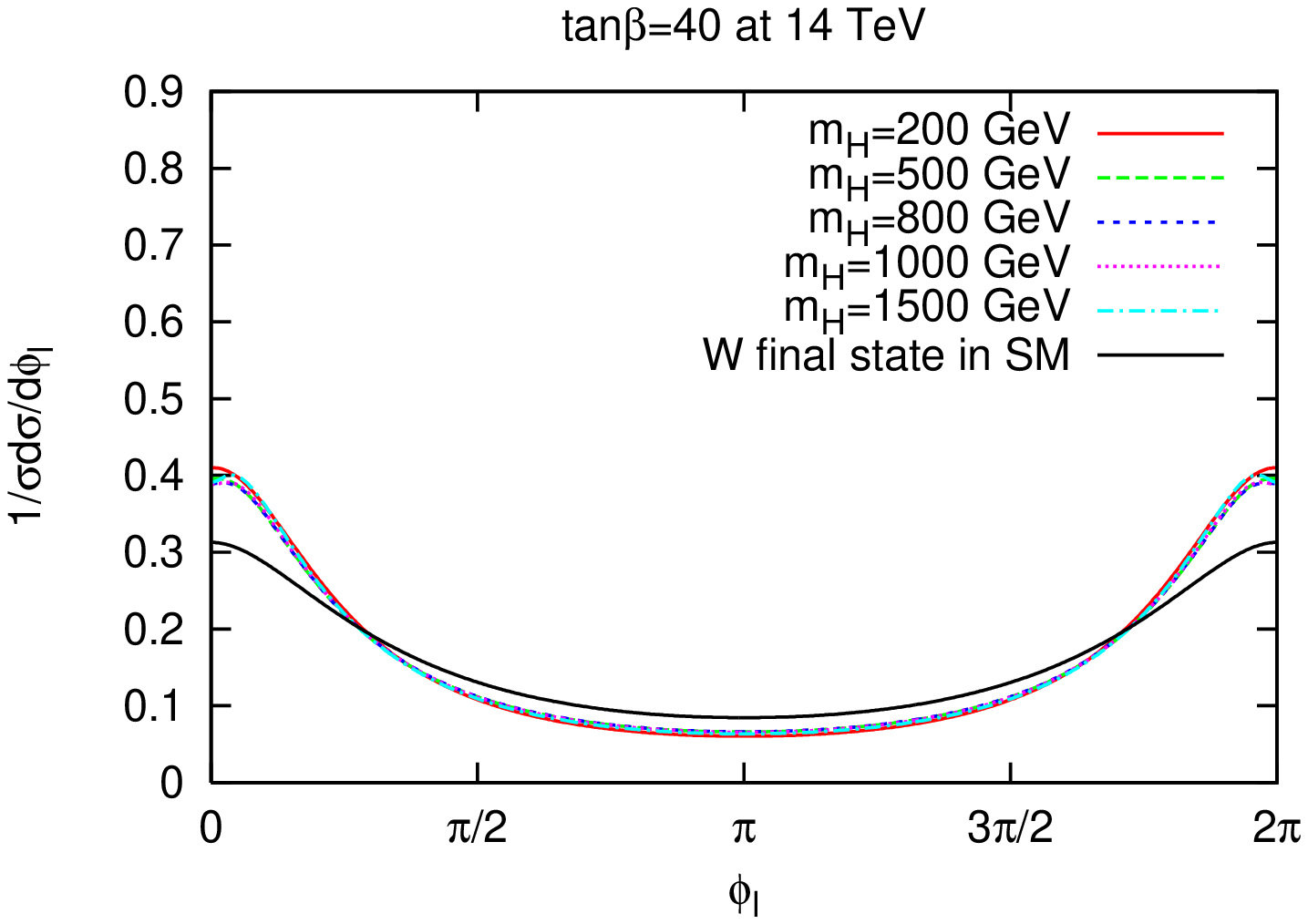} 
\caption{\sl The normalized lepton azimuthal distribution for $\tan \beta=5$ (left) and $\tan\beta=40$ (right) for various charged Higgs masses at a cm energy of 14 TeV.} 
\label{phidis14TeV}
\end{center}
\end{figure}

With this choice of frame, the normalized lepton azimuthal distributions for $\sqrt{s}=7$ TeV is shown in Fig. \ref{phidis7TeV} for small and large values of $\tan \beta$, for various $M_{H^-}$ values. The corresponding plots for a cm energy of 14 TeV is shown in Fig. \ref{phidis14TeV}. The $\phi_\ell$ distribution for $tW^-$ production in the SM is also shown for comparison. 

The $\phi_\ell$ distributions for other values of $\tan \beta$ and $M_{H^-}$ have a similar profile, with a peak at $\phi_\ell=0$ and $2 \pi$. The $\phi_\ell$ distribution depends on both kinematic and top polarization effects and the factors which influence its shape can be understood as follows. According to Eqn. (\ref{topdecaywidth}), the decay lepton is emitted preferentially along the top spin direction in the top rest frame, with $\kappa_f=1$. To obtain the distribution in the laboratory frame we use the following relation between the angle $\theta_{\ell}^*$ between the top spin and decay lepton in the top rest frame and the angle $\theta_{t \ell}$ between the top and lepton in the laboratory frame:
\begin{equation}
 \cos \theta_{\ell}^*=\frac{\cos \theta_{t \ell}-\beta}{1-\beta \cos \theta_{t \ell}}
\label{boost}
\end{equation}
where
\begin{equation}
 \cos \theta_{t \ell}=\cos \theta_t \cos \theta_{\ell}+\sin \theta_t \sin \theta_{\ell} \cos \phi_{\ell}.
\label{costhetatl}
\end{equation}
Using the above relations, the laboratory frame angular distribution of the lepton becomes
\beq\label{thetatldist}  \displaystyle
\frac{1}{\Gamma_{\ell}}\frac{d\Gamma_{\ell}}{d\cos\theta_{t\ell}} = \displaystyle \frac{1}{2}
(1-\beta^2)(1 - P_t \beta)\frac{1 + \frac{P_t - \beta}{1 - P_t
\beta} \cos\theta_{t\ell}}{(1- \beta
\cos\theta_{t\ell})^3},
\eeq
where $\beta=\sqrt{1-m_t^2/E_t^2}$ is the top velocity in the parton cm frame. We notice that the distribution (\ref{thetatldist}) peaks for large $\cos \theta_{t \ell}$, since it occurs in the denominator and hence from Eq. (\ref{costhetatl}) for small $\phi_{\ell}$. Thus, the boost to the laboratory frame produces a collimating effect along the direction of the top momentum, which gets translated to a peaking at $\phi_{\ell}=0$. 

\begin{figure}
 \begin{center}
 \includegraphics[angle=270,width=3.2in]{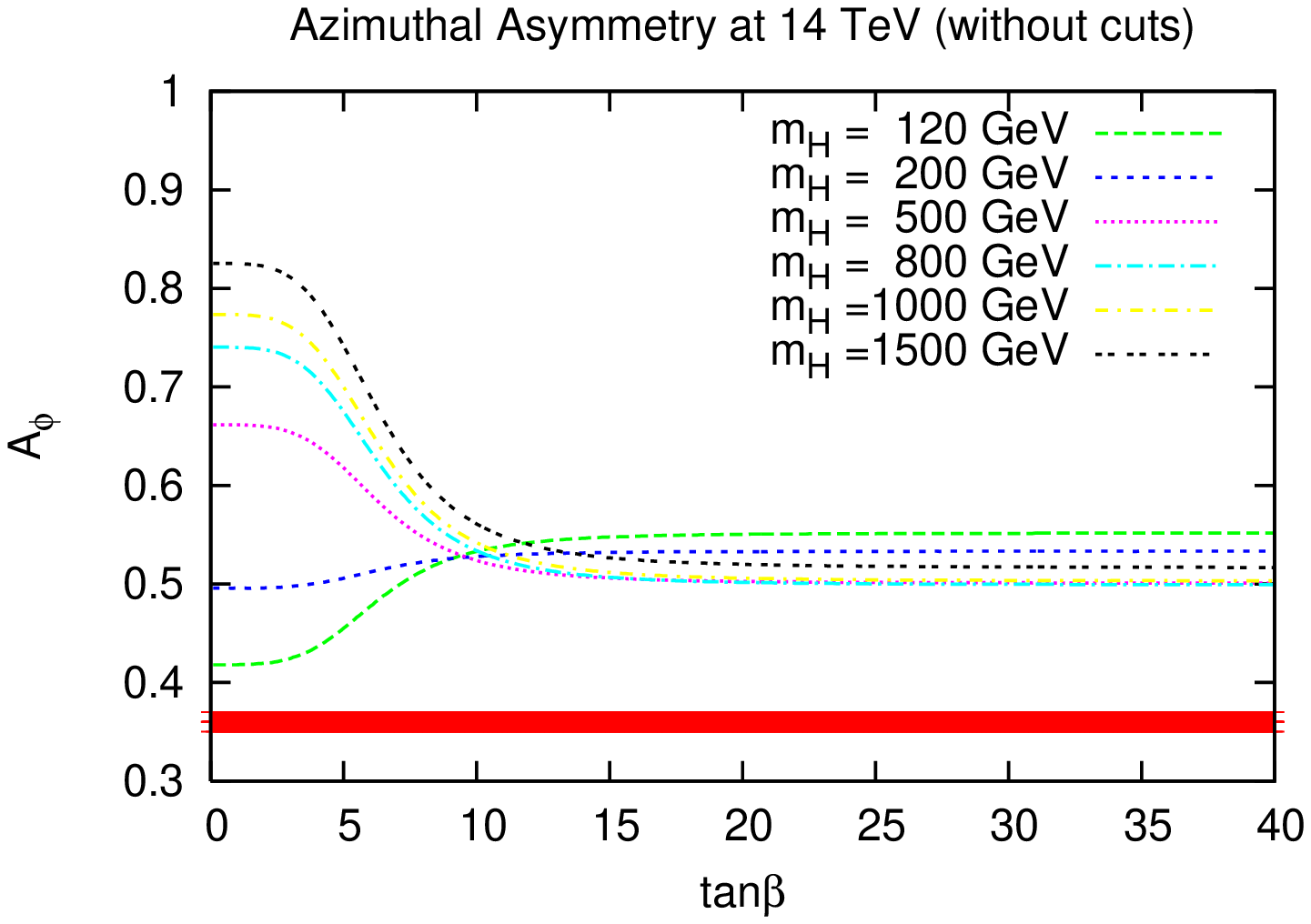}
  \includegraphics[angle=270,width=3.2in]{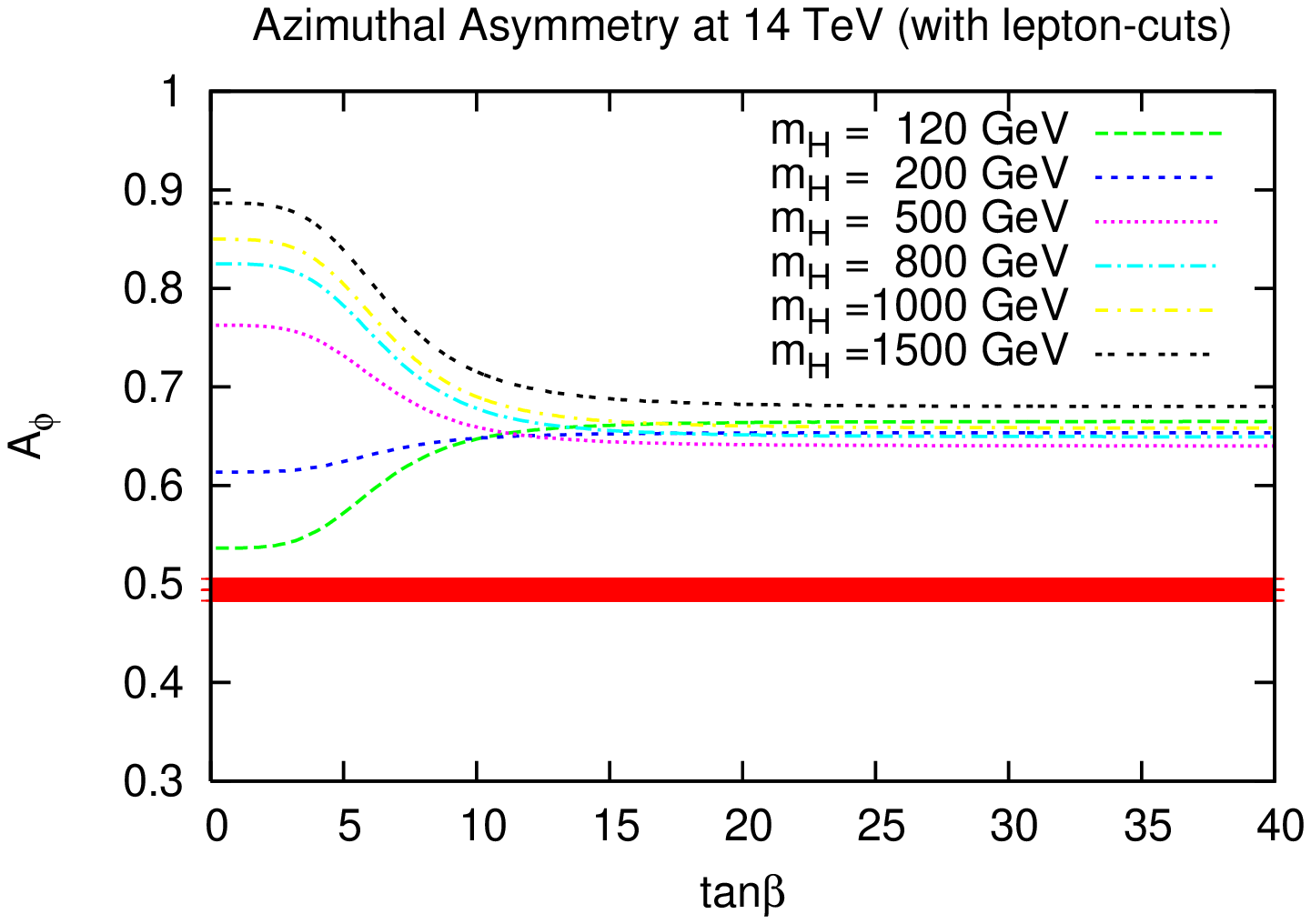} 
 \caption{\sl $A_{\phi}$ as a function of $\tan\beta$ and different charged Higgs masses at $\sqrt{s}=14$ TeV without lepton cuts (left) and with cuts (right). The red band corresponds to the azimuthal asymmetry for $tW$ production in the SM with a $2 \sigma$ error interval.}
 \label{aziasy14lepton}
 \end{center}
 \end{figure}

We notice that the curves are separated at the peaks for different $M_{H^-}$ values and are very different from the $tW$ case in the SM. As in \cite{Godbole:2010kr, Godbole:2006tq,leshouch,Godbole:2009dp}, we can quantify this difference by defining a normalized azimuthal asymmetry for the lepton as
\begin{equation}
 A_{\phi}=\frac{\sigma(\cos \phi_\ell >0)-\sigma(\cos \phi_\ell<0)}{\sigma(\cos \phi_\ell >0)+\sigma(\cos \phi_\ell<0)},
\label{aziasy}
\end{equation}
where the denominator is the total cross section. A plot for $A_{\phi}$ as a function of $\tan \beta$ with and without cuts on the lepton momenta are shown in Fig. \ref{aziasy14lepton} for a cm energy of 14 TeV. We have used the following rapidity and transverse momentum acceptance cuts on the decay lepton: $|\eta|<2.5,\,p_{T}^\ell>20$ GeV. Also shown is the SM value for $A_{\phi}$ for $tW$ production with a $2 \sigma$ error band. 

 The lepton cuts only mildly increase the value of $A_{\phi}$ for the charged Higgs case and the value for $tW$ production in the SM is also enhanced from about 0.35 without cuts to about 0.5 with cuts, as can been seen from Fig. \ref{aziasy14lepton}. The azimuthal asymmetry also shows considerable variation, as a function of $\tan \beta$, roughly in the range $3 \lesssim \tan \beta \lesssim 15$ and becomes flat for values outside this range and almost independent of $M_{H^-}$. From Fig. \ref{poltanb}, we see that this is the same range of $\tan \beta$ for which the polarization $P_t$ shows variation, becoming constant for roughly $\tan\beta > 15$; thus, the azimuthal asymmetry follows the same trends as the top  polarization. If the mass of the charged Higgs is known, from a measurement of $A_{\phi}$ it would be easier to determine $\tan\beta$ if it lies within this range.

\begin{figure}[h]
\begin{center}
 \includegraphics[angle=270,width=3.2in]{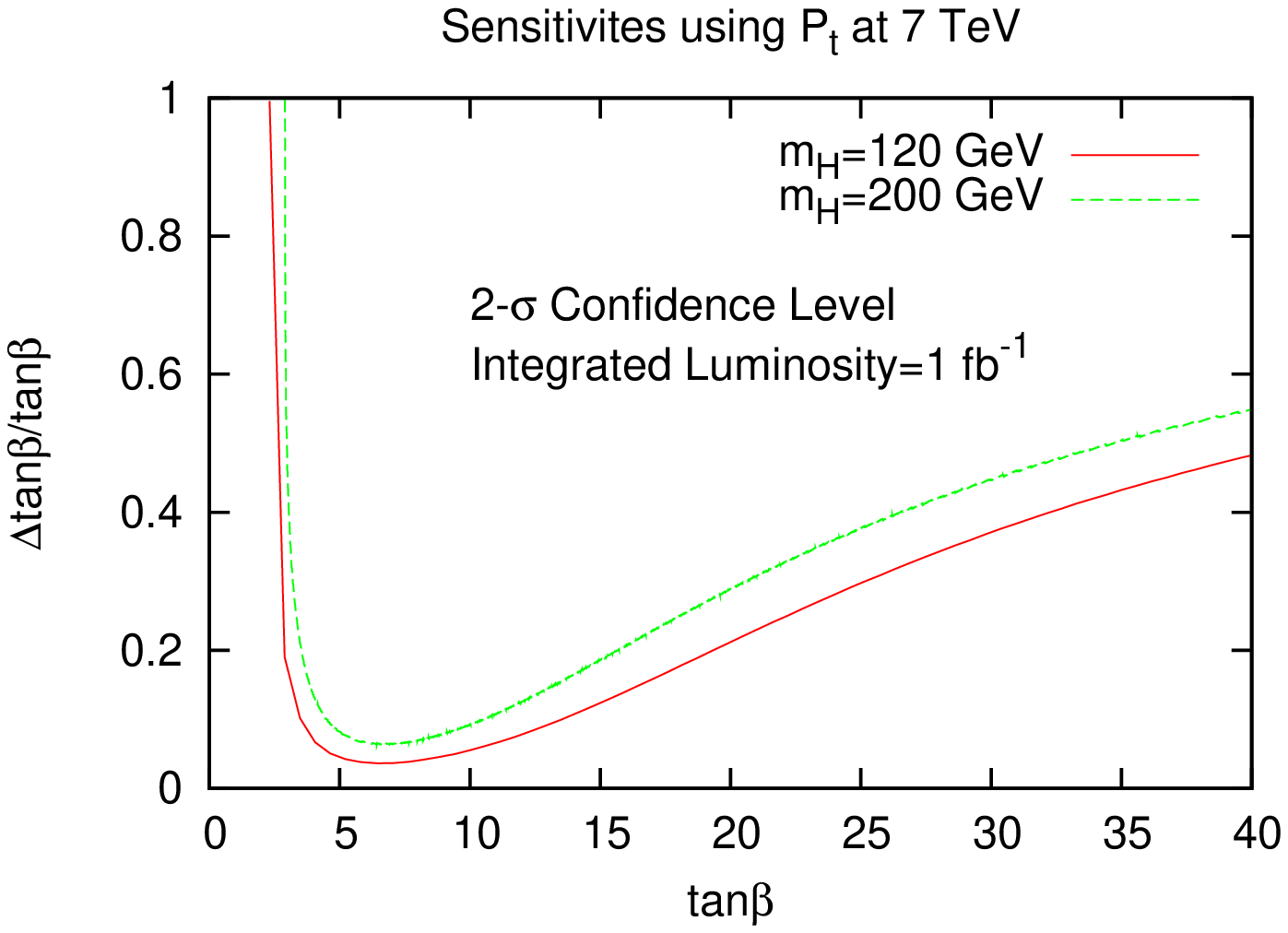}
  \includegraphics[angle=270,width=3.2in]{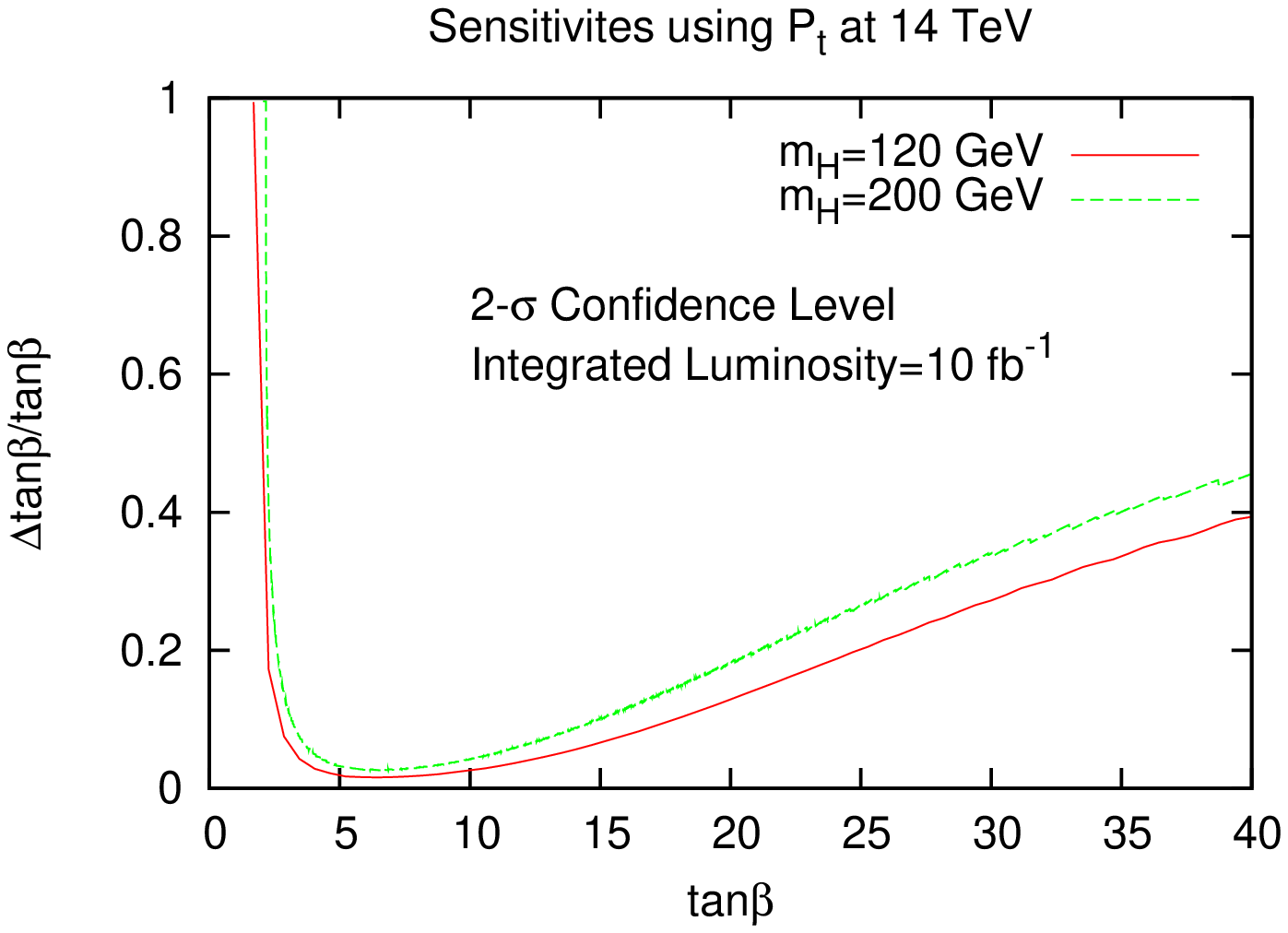} 
 \caption{\sl The fractional accuracy of $\tan\beta$ at 2$\sigma$ CL as a function of $\tan\beta$ for $\sqrt{s}=7$ TeV (left) and 14 TeV (right) using the polarization $P_t$, with $\int \mathcal{L} dt =1$ fb$^{-1}$ and $10$ fb$^{-1}$ respectively.}
\label{tanbPtsensitivity}
\end{center}
\end{figure}

 We now investigate the accuracy to which one can determine $\tan\beta$ from the top polarization, $P_t$, and the azimuthal asymmetry, $A_{\phi}$. The accuracy of the determination of parameter $\tan\beta$ at $\tan\beta_0$, from the measurement of an observable $O(\tan\beta)$, is $\Delta \tan\beta$ if $|O(\tan\beta)-O(\tan\beta_0)|<\Delta O(\tan\beta_0)$ for $|\tan\beta_0-\tan\beta|<\Delta \tan\beta$, where $\Delta O(\tan\beta_0)$ is the statistical fluctuation in $O$ at an integrated luminosity $\mathcal L$. The corresponding fractional accuracy is then $\Delta \tan\beta/\tan\beta_0$. For top-polarization, $P_t$ and azimuthal asymmetry, $A_{\phi}$, the statistical fluctuations at a level of confidence $f$ are given by $\Delta O=f/\sqrt{\mathcal L \sigma}\times \sqrt{1-O^2}$, where $O$ denotes $P_t$ or $A_{\phi}$.

 In Fig. \ref{tanbPtsensitivity}, we show the fractional accuracy $\Delta \tan\beta/\tan\beta$
in the determination of the coupling $\tan\beta$ from the polarization
$P_t$ at 2$\sigma$ confidence level (CL). We choose, for illustration, charged Higgs masses of 120 and 200 GeV and an integrated luminosity of 1 fb$^{-1}$ and 10 fb$^{-1}$ for $\sqrt{s}=7$ and 14 TeV respectively. 
We use, for convenience, the criterion $\Delta\tan\beta/\tan\beta < 0.3$
for an accurate determination of $\tan\beta$ since this corresponds to a
relative accuracy of about 1\% in the determination of physical
quantities, which are proportional to the square of the couplings.

 Then, we see that at $\sqrt{s}=7$ TeV, $\tan\beta$ can be determined accurately for values between roughly 3 and 25 for $M_{H^-}=120$ GeV and between 3 and 20 for $M_{H^-}=200$ GeV. The corresponding range for $\tan\beta$ determination for the LHC running at 14 TeV are 3 to 30 for $M_{H^-}=120$ GeV and 3 to 25 for $M_{H^-}=200$ GeV. For larger $\tan \beta$ (and even for very low $\tan\beta$) the sensitivity worsens since the $P_t$ curves become flat and do not show much variation as a function of $\tan\beta$, as can be seen from Fig. \ref{poltanb}. One can, of course, choose a different value for $\Delta \tan\beta/\tan\beta$ as a measure of $\tan \beta$ accuracy in which case the corresponding limits on $\tan \beta$ will be different as can be read from the plots.

\begin{figure}[h]
\begin{center}
 \includegraphics[angle=270,width=3.2in]{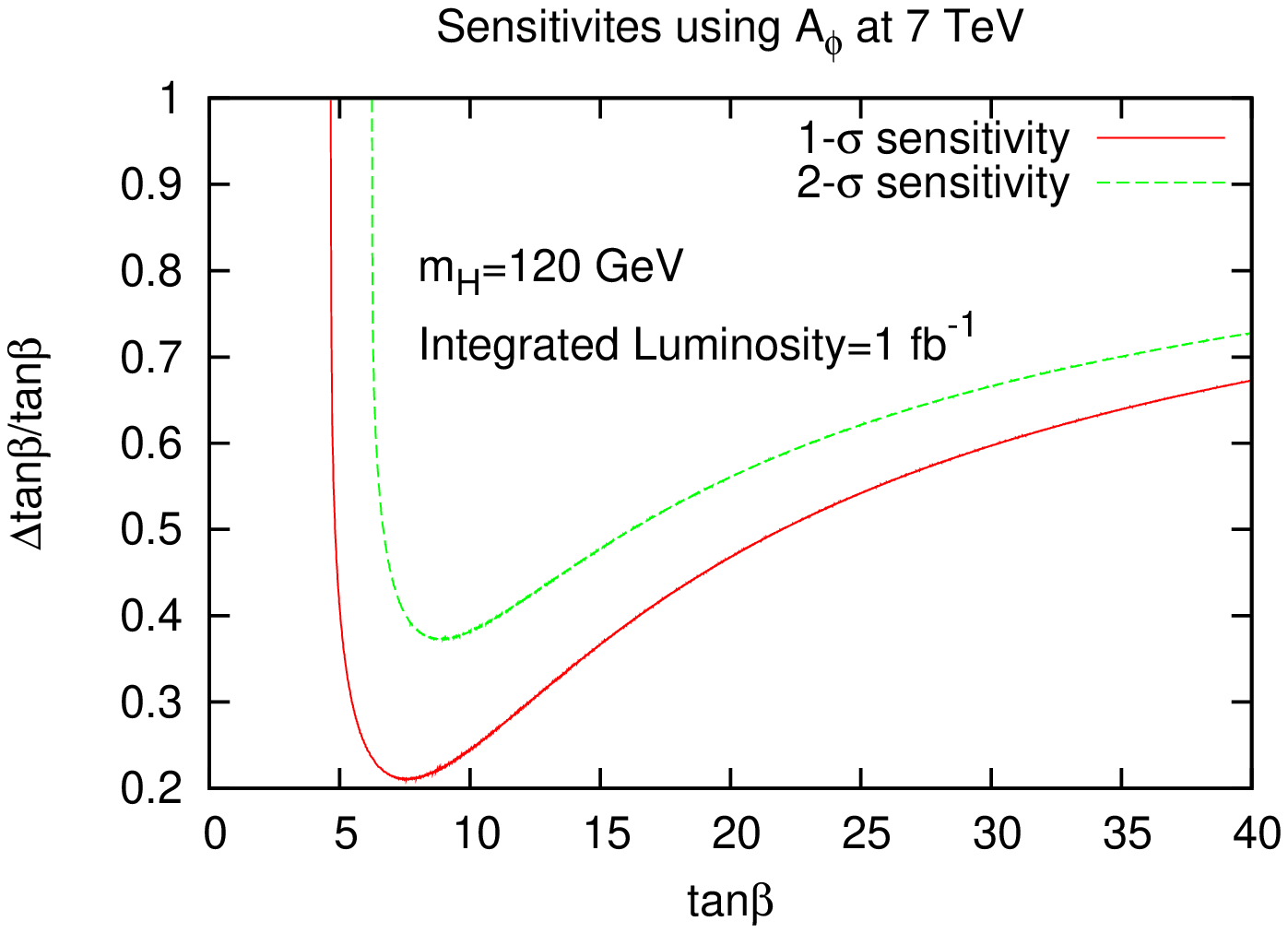} 
 \includegraphics[angle=270,width=3.2in]{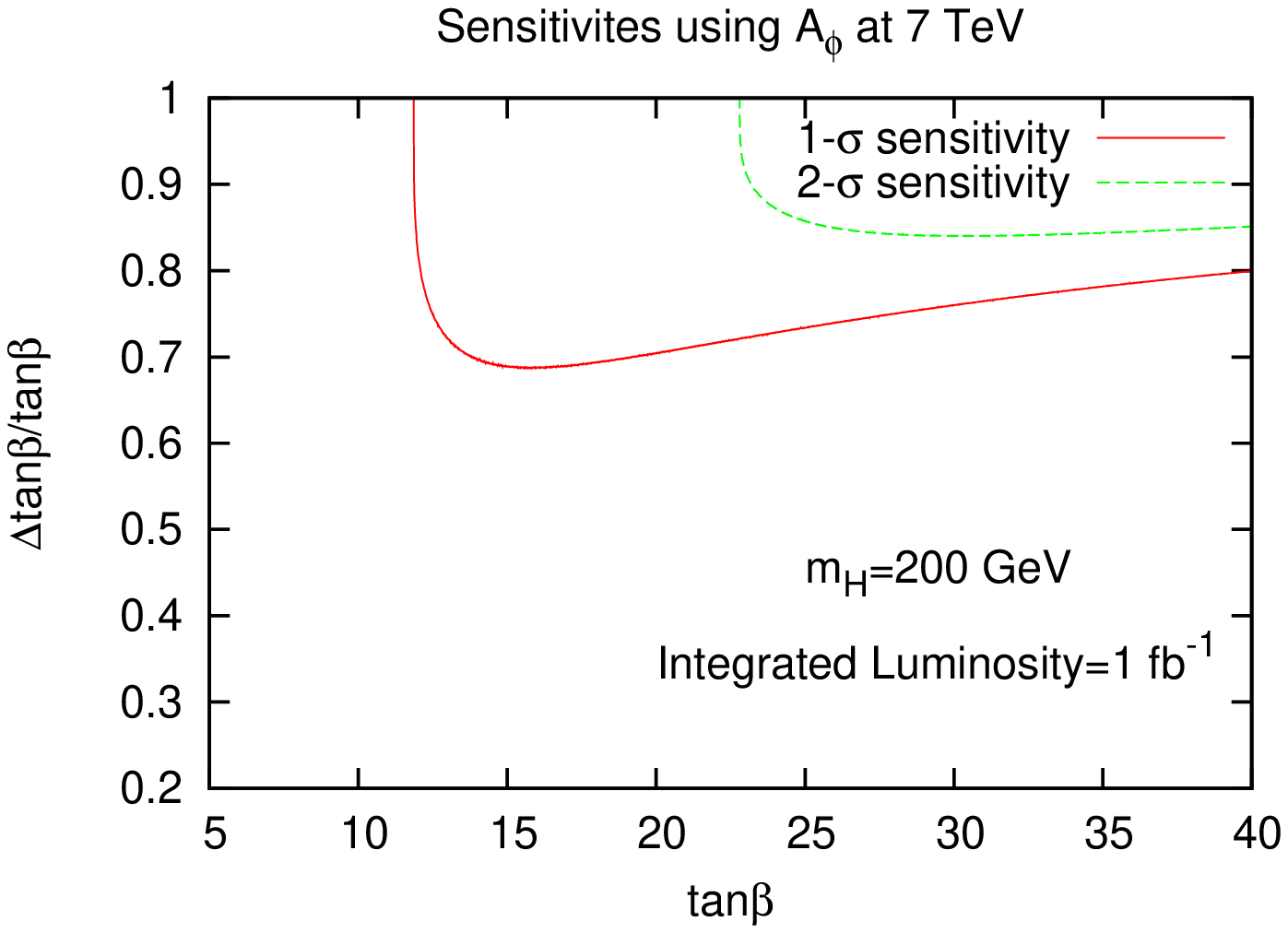}
 \caption{\sl The fractional accuracy of $\tan\beta$ as a function of $\tan\beta$ for $\sqrt{s}=7$ TeV using the azimuthal asymmetry $A_{\phi}$ for $M_{H^-}=120$ GeV (left) and $M_{H^-}=200$ GeV (right).}
\label{tanbAziAsysensitivity7TeV}
\end{center}
\end{figure}

\begin{figure}[h]
\begin{center}
 \includegraphics[angle=270,width=3.2in]{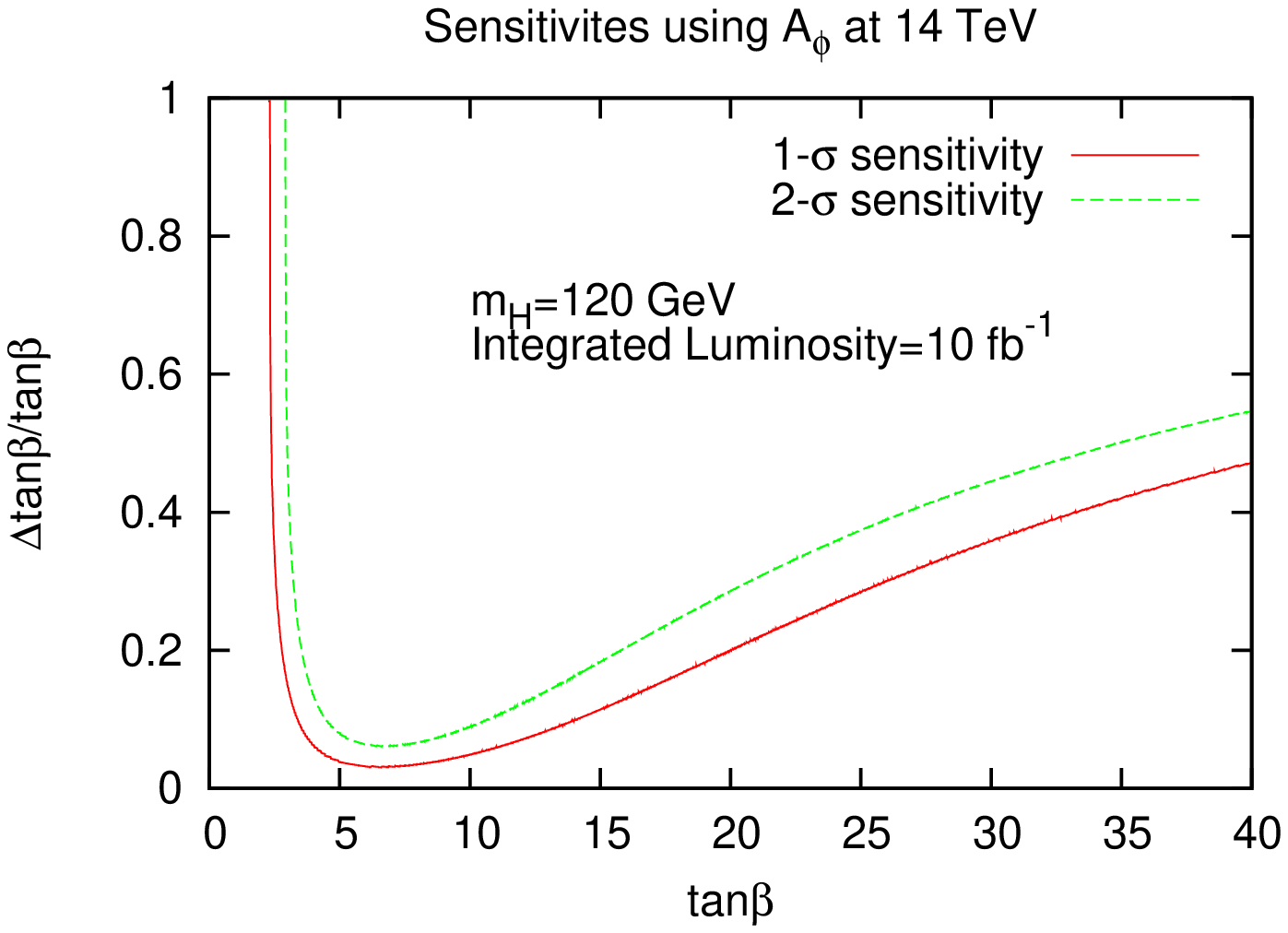} 
 \includegraphics[angle=270,width=3.2in]{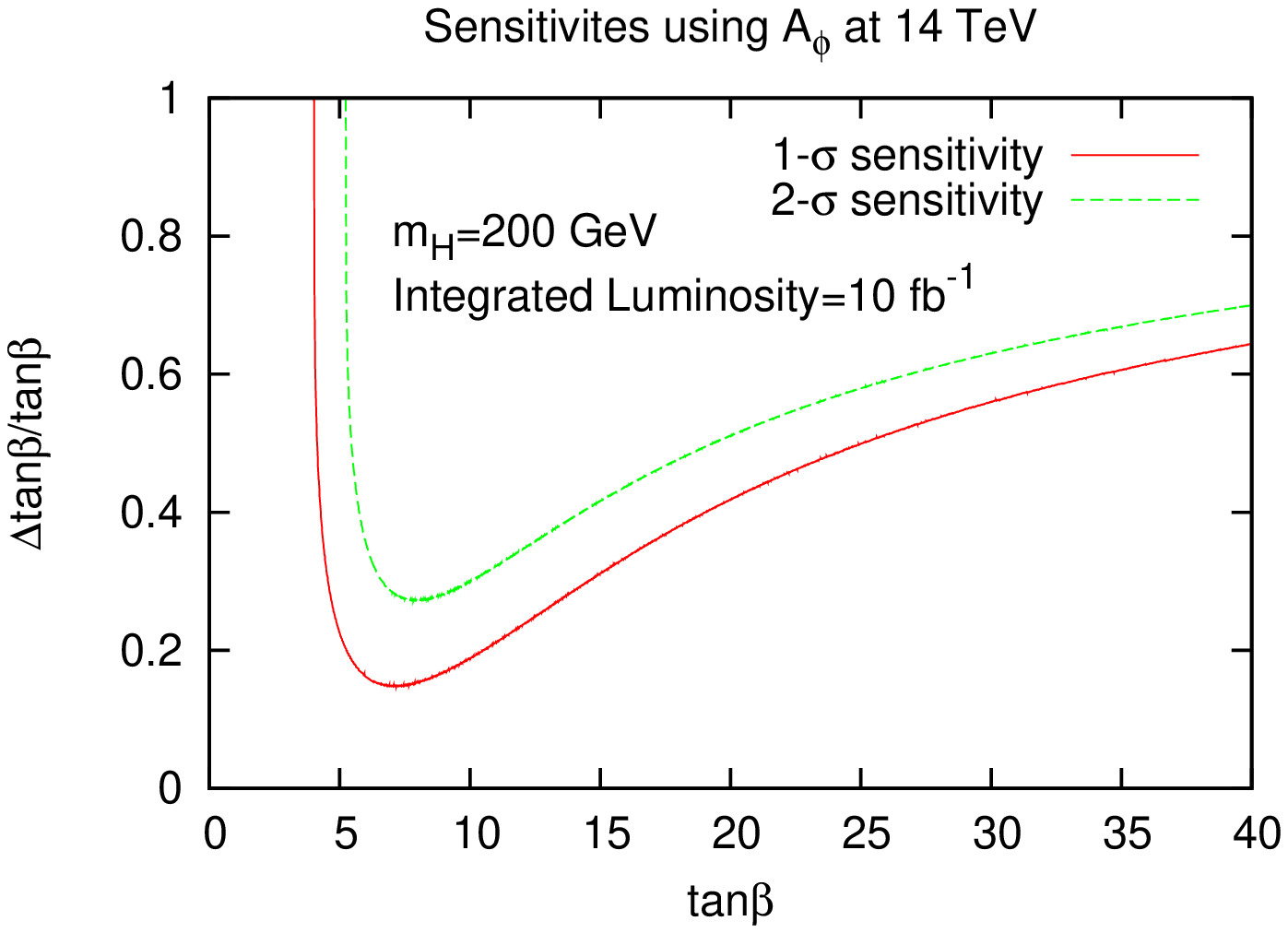}
 \caption{\sl The fractional accuracy of $\tan\beta$ as a function of $\tan\beta$ for $\sqrt{s}=14$ TeV using the azimuthal asymmetry $A_{\phi}$ for $M_{H^-}=120$ GeV (left) and $M_{H^-}=200$ GeV (right).}
\label{tanbAziAsysensitivity14TeV}
\end{center}
\end{figure}

We now consider the accuracy to which $\tan \beta$ can be determined from the more conveniently measurable azimuthal asymmetry. Plots of the fractional accuracy for this case are shown in Fig. \ref{tanbAziAsysensitivity7TeV} and Fig. \ref{tanbAziAsysensitivity14TeV} for the cases of $\sqrt{s}=7$ TeV and 14 TeV respectively and with the indicated charged Higgs masses and luminosities. 
If we use the same criterion for $\tan\beta$ accuracy as before, $\Delta\tan\beta/\tan\beta < 0.3$, we notice that for a cm energy of 7 TeV and an integrated luminosity of 1 fb$^{-1}$, the azimuthal asymmetry is not a very sensitive measure of $\tan\beta$. For the lower charged Higgs mass of 120 GeV, and at a 1$\sigma$ CL, $\tan\beta$ can be probed roughly in the range 6 to 12; the sensitivity worsens for larger charged Higgs masses or CL's. The top polarization is better probe of $\tan\beta$ than the azimuthal asymmetry.
  However, this is due to the fact that in constructing the asymmetry only the semi-leptonic decay modes of the top have been considered, which reduces the cross section by a factor of 3. The sensitivities are considerably enhanced if we include all decay channels of the top. But it must be remembered that using any decay product of the top other than $\ell^+$ and $\bar{d}$ to construct the azimuthal asymmetry will make $A_{\phi}$ dependent on new physics in the $tbW$ vertex. For the LHC running at $\sqrt{s}=14$ TeV, $A_{\phi}$ is a more sensitive measure of $\tan\beta$ compared to the 7 TeV case, at least for the lower charged Higgs mass of 120 GeV. For this case $\tan\beta$ can be probed in the range 3 to 25 at the $1\sigma$ CL and between 3 and 20 at a $2\sigma$ CL. For $M_{H^-}=200$ GeV, $A_{\phi}$ is sensitive to $\tan\beta$ only at the $1\sigma$ CL for a smaller range of 5 to 15.

As is to be expected, $\tan\beta$ can be determined to a higher accuracy and for a larger range using the top polarization $P_t$, compared to the azimuthal asymmetry constructed from the decay lepton; the restriction to semi-leptonic decay modes of the top further reduces the sensitivity to $A_{\phi}$. However, it is interesting to note that the profile of the plot of $\Delta \tan \beta / \tan \beta$ vs $\tan \beta$ computed by using $A_{\phi}$, shown in Fig. \ref{tanbAziAsysensitivity7TeV} and \ref{tanbAziAsysensitivity14TeV}, is similar to that obtained by using the polarization $P_t$, shown in Fig. \ref{tanbPtsensitivity}. $A_{\phi}$ follows the change in $P_t$ as a function of the coupling $\tan\beta$ and is thus a faithful probe of the top polarization itself. At least for $\sqrt{s}=14$ TeV and $M_{H^-}=120$ GeV, the range in which $\tan\beta$ can be probed accurately using $A_{\phi}$ or $P_t$ is roughly similar for both variables.

Thus, the azimuthal asymmetry can be a convenient and sensitive probe of both the top polarization and the coupling parameter $\tan \beta$ in the THDM, at least in the regions of parameter space mentioned above.

It is worthwhile to comment on the dominant backgrounds to our signal process $gb\rightarrow tH^-\rightarrow t\bar{t}b$.  
When $M_{H^-}>m_t + m_b$, we require the top to decay semi-leptonically and the anti-top to decay 
hadronically to trigger on the charged Higgs signal, as well as for the purpose 
of reconstruction of the top quarks and the charged Higgs. The complete final 
state therefore consists of 3 $b$ jets + 2 light jets + 1 lepton + missing energy. 
The main background for this signal would come from next-to-leading order NLO QCD processes,
which are (a) $gg\rightarrow t\bar{t}b\bar{b}$, (b) $gb\rightarrow t\bar{t}b$, and (c) $gg\rightarrow t\bar{t}g$, 
where in the first case, one of the b jets is missed and in the last case the gluon jet is mis-tagged as a $b$ jet (with probability of around 1 \%). Refs.\cite{Moretti:1999bw, CMS, 
Assamagan:2004tt} have investigated the charged-Higgs signal in this process in 
great detail for the LHC with triple $b$-tagging. They have used  kinematical 
cuts of $p_T>30$ GeV and $|\eta|<2.5$ for all jets and assume $b$-tagging 
efficiency of 40\% in their analysis. The conclusion from their analysis for
30 fb$^{-1}$ of accumulated data is that there are enough number of events for 
charged Higgs discovery in this channel at the 5-$\sigma$ level upto a mass of 
600 GeV for very large values of $\tan\beta$ ($>25$) and very small values of 
$\tan\beta$ ($<5$). We can expect better visibility for the charged Higgs when the $b$-tagging efficiency increases in future. 
Backgrounds from weak processes like $tW+X,~ b \bar{b}+X$ and $W + 2j$ would be 
suppressed because we choose the signal to consist of 3 $b$ jets and an isolated lepton.

When $M_{H^-}< m_t +m_b$, the dominant decay of the $H^-$ is into $\tau
+\bar \nu_\tau$.  Our signal in
this will be $gb\rightarrow tH^-\rightarrow t\tau^- \bar \nu_\tau
\rightarrow b \ell^+ \nu_\ell \tau^-\bar \nu_\tau$. 
For this final state of
$b$ + lepton + $\tau$ + missing energy, the background now comes
from the processes of $t\bar t$ production with the $\bar t$ decaying into a
$\tau$ and $tW^-$ production with $W^-$ decaying into a $\tau$. In both
these cases, since the $\tau$ comes from $W^-$ decay, $\tau$
polarization can be used to suppress the background \cite{Roy:2004az}.
While the presence
of two neutrinos in the final state would seem to 
make it impossible to reconstruct
the top production plane needed for our analysis, we are helped by the
fact that the $tH^-$ events are produced close to the threshold because
of the sharp peaking of the initial-state partons at low $x$. Thus it is
a reasonable approximation to treat the top quark and the charged Higgs
as at rest, enabling approximate determination of the energy and momenta
of both neutrinos on an event-by-event basis. 

The NLO QCD corrections to the process $gb\rightarrow tH^-$ have been studied in
 Refs.\cite{Plehn:2002vy,Zhu:2001nt} and next-to-next-to-leading-order (NNLO) 
soft gluon corrections have been evaluated in
Ref.\cite{Kidonakis:2004ib}. These corrections are shown to be
substantial, upto 85 \% of the LO cross section for large Higgs masses. 
It has been also shown that the K-factor in this process is proportional to the 
mass of charged Higgs. Since QCD corrections are model independent, one
can use the K-factor appropriately in the analysis to rescale the LO result to 
the NLO order. The normalized differential cross sections and the
asymmetries we calculate would be insensitive to the higher order
corrections. We have not used any K-factor in our analysis. 
Including NLO QCD corrections through the naive use of K-factor would 
increase our signal cross section by a factor of 1.5-1.85 depending upon the 
charged Higgs mass and hence sensitivity to the parameters would increase. 

The complete NLO EW calculations for the process $gb\rightarrow tH^-$
have been done in Ref.\cite{Beccaria:2009my} for type II 2HDM. 
They have reported that the NLO EW correction to the total cross section is very
 mild. It varies from less than 1\% for low values of $\tan\beta$ to less than 
4\% for higher values of $\tan\beta$. The effects of NLO EW corrections to      
observables like top polarization, normalized angular distributions and angular 
asymmetries are expected to be small. For example, in Ref. 
\cite{Beccaria:2004xk}, it has been shown that NLO EW supersymmetric effects on 
top polarization is almost zero for all values of charged Higgs masses and all 
values of $\tan\beta$ except for $\tan\beta\approx 10$, for which correction is 
around -1\% to -3\%. 

Any NLO corrections to top decay will not affect our analysis of charged lepton 
angular distributions and asymmetries as it has been proven that charged lepton 
angular distributions are independent of any corrections to form factors
in top decay. There can also be NLO corrections from non-factorizable diagrams. 
However, this analysis has not been done in the literature so far and it would 
be interesting to see the effect of these non-factorizable diagrams to our 
analysis which is beyond the scope of this work. 

\section{Summary}
We have studied the issue of using the polarization of the top quark produced in association with a charged Higgs in the type II THDM or SUSY models as a probe of the coupling parameter $\tan \beta$ occurring in such models. Since the top decays before it has the time to hadronize, its polarization, reflected in the angular distribution of its decay products, can be a probe of new physics underlying its production. We have derived analytic expressions for  left and right polarized $tH^-$ production (and the off-diagonal elements as well in the spin density matrix). Essentially because of the scalar-pseudoscalar coupling (\ref{tbhcoupling}) of the $tbH^-$ vertex, compared to the vector-axial vector couplings of the top in the SM, the resulting polarizations are vastly different from that expected in the SM and are sensitively dependent on the charged Higgs mass and $\tan \beta$, as shown in Figs. \ref{poltanb} and \ref{polmh}, where we considered both the cm energies of 7 and 14 TeV at which the LHC is planned to run. The degree of longitudinal top polarization can be as large as 0.3 to 0.4 (for a charged Higgs mass of 120 GeV and for $\tan\beta$ values less than 5 and greater than 10), compared to the SM values of $-0.25$ for $tW$ production or $\mathcal{O}(-10^{-4})$ for $t\bar{t}$ production. Characteristic of the $tbH^-$ coupling in the THDM, the $2 \to 2$ top production cross sections are minimized and the polarizations vanish and change sign as a function of $\tan \beta$ at $\tan \beta=\sqrt{\frac{m_t}{m_b}}$. 

We then investigated to what extent top polarization is reflected in the angular distribution of the decay lepton in the process $t\to bW^+ \to b \nu_{\ell} \ell^+$, with inclusive decay of the $b$ and $H^-$. Since it is known that the laboratory frame angular distributions of the charged lepton in top decay depends only on the top production process and are independent of new physics in the $tbW$ vertex, we considered the azimuthal distribution of the lepton from top decay, $A_{\phi}$, as a probe of new physics in its production  (we find the polar distribution of the lepton in the THDM insensitive to $\tan\beta$ and the charged Higgs mass and almost identical to $tW$ production in the SM). $A_{\phi}$ is sensitive to $\tan \beta$ values roughly in the range $3 \lesssim \tan \beta \lesssim 15$, for different charged Higgs masses considered and becomes constant for larger $\tan\beta$ values. This is the same range in which the top polarization shows variation as a function of $\tan\beta$; $A_{\phi}$ thus captures the dependence of $P_t$ on $\tan\beta$. If the charged Higgs mass is already known, a measurement of $A_{\phi}$ can help measure $\tan \beta$ if it lies in the above range. 

We also computed the fractional accuracy to which $\tan \beta$ can be measured, as a function of $\tan \beta$, from the top polarization $P_t$ and a measurement of the azimuthal asymmetry $A_{\phi}$. Using the criterion that $\Delta \tan \beta / \tan \beta <0.3$ for an accurate determination of $\tan \beta$, we find that $P_t$ can help determine $\tan \beta$ lying in the range between 3 and 25 for a cm energy of 7 TeV and between 3 and 30 for the 14 TeV case, at a $2 \sigma$ CL for $M_{H^-}=120$ GeV; the range is only slightly smaller for a larger $M_{H^-}$ of 200 GeV. While the azimuthal asymmetry is not very sensitive to an accurate measurement of $\tan\beta$ for the LHC running at 7 TeV, we find that at 14 TeV one can use the azimuthal asymmetry to probe $\tan\beta$ up to 25 at a $1 \sigma$ CL and for $M_{H^-}=120$ GeV; for $M_{H^-}=200$ GeV the corresponding range is 5 to 15. Including both leptonic and hadronic decay modes of the top is expected to increase the sensitivity of the azimuthal asymmetry to $\tan\beta$; however, this renders the asymmetry sensitive to new physics in the $tbW$ decay vertex, apart from new physics in top production.

The sensitivity plot for $\tan \beta$ determination using $A_{\phi}$ follows roughly the one obtained by using $P_t$. Thus, the azimuthal asymmetry of the decay lepton can be a convenient and accurate probe of the top polarization and the coupling parameter $\tan \beta$ of the THDM or SUSY models for the LHC running at $\sqrt{s}=14$ TeV and for smaller charged Higgs masses.

\section{Acknowledgements}
K.H and K.R gratefully acknowledge support from the Academy of Finland (Project No. 115032). S.K.R is supported by US Department of Energy, Grant Number DE-FG02-04ER41306. S.D.R. thanks Helsinki Institute
of Physics and the University of Helsinki for hospitality during the period when this work was
completed.



\end{document}